# Security Analysis of the Silver Bullet Technique for RowHammer Prevention

21 January 2021

A. Giray Yağlıkçı, SAFARI Research Group, ETH Zürich

Jeremie S. Kim, SAFARI Research Group, ETH Zürich

Fabrice Devaux, UPMEM

Onur Mutlu, SAFARI Research Group, ETH Zürich

SAFARI

SAFARI Research Group

up mem

# Table of Contents

## Executive Summary

The purpose of this document is to study the security properties of the Silver Bullet[1] algorithm against worst-case RowHammer attacks. We mathematically demonstrate that Silver Bullet, when properly configured and implemented in a DRAM chip, can securely prevent RowHammer attacks. The demonstration focuses on the most representative implementation of Silver Bullet, the patent claiming many implementation possibilities not covered in this demonstration. Our study concludes that Silver Bullet is a promising RowHammer prevention mechanism that can be configured to operate securely against RowHammer attacks at various efficiency-area tradeoff points, supporting relatively small hammer count values (e.g., 1000) and Silver Bullet table sizes (e.g., 1.06KB).

## 1. Introduction

RowHammer [8] is a circuit-level DRAM vulnerability, where repeatedly activating (i.e., opening and closing) a DRAM row can cause bit flips in nearby rows due to circuit-level disturbance effects and accelerated capacitor charge leakage [2,3,8,9,11-13,19,20]. The vulnerability of a DRAM chip to a RowHammer attack is measured by the unsafe hammer count ($UHC_{DRAM}$[2]) or the minimum number of activations required to cause a RowHammer bit flip in the chip.

Silver Bullet [10] is a new mechanism that mitigates RowHammer bit flips by effectively increasing the refresh rate for localized regions of DRAM that have high activation rates. The purpose of this document is to rigorously analyze the security guarantees provided by the Silver Bullet technique. We demonstrate via mathematical proof that, if its parameters are set correctly, Silver Bullet can provide a provably-secure solution to RowHammer bit flips.

Silver Bullet effectively increases the refresh rate by counting DRAM activations that are issued to each subbank (i.e., a group of physically-contiguous rows) and issuing a preventive DRAM refresh to one row in the subbank each time the subbank's activation counter ($FRAC$) reaches a predefined value ($PARAMD$, which we denote as $D$ for simplicity). The row to be refreshed within a subbank is identified by a pointer that cycles through each row of the subbank.

In this document, we analyze the security guarantees of Silver Bullet against RowHammer bit flips. We calculate Silver Bullet's tolerable hammer count (i.e., THC; the minimum $UHC$ value of a DRAM chip in which Silver Bullet can effectively prevent RowHammer bit flips) for various configuration parameters and implementations. We derive the worst-case DRAM access pattern against a system protected by Silver Bullet and present the tradeoffs of $THC$ and various Silver Bullet configuration parameters and DRAM device characteristics.

The remainder of this document is organized as follows. Section 2 provides an overview of the Silver Bullet mechanism. Section 3 states our assumptions in its mathematical security proof. In Section 4, we discuss the attacker model and Silver Bullet features that enable an attacker to craft the worst-case RowHammer attack. Section 5 provides a mathematical representation for any RowHammer attack, i.e., any attack that aims to increase the hammer count to a value equal to or greater than $UHC_{DRAM}$. Section 6 analyzes the worst-case access patterns and calculates the maximum number of activations that an attacker can issue (before Silver Bullet performs preventive refresh to all rows in a subbank that is targeted by the attacker). Section 7 calculates the table size in terms of Silver Bullet configuration parameters. Based on this analysis, we define the set of constraints that an implementation of Silver Bullet should satisfy to ensure RowHammer-safe operation in Section 8. In Section 9, we provide an overview of the hammer count values Silver Bullet can securely tolerate (i.e., $THC$ values) with different configuration of its parameters and the resulting overhead in terms of additional preventive refreshes it performs, given the set of security constraints we define in Section 8. Section 10 analyzes the sensitivity of Silver Bullet's security guarantees to key Silver Bullet design parameters and target device characteristics. Section 11 reiterates our key findings and conclusion.

[1] Silver Bullet refers to an UPMEM patent [10] which can be licensed under RAND terms (Raisonable And Non Discriminatory).
[2] This value is called MAC [1], RowHammer threshold [2,8,14,15], and $HC_{first}$ [3] in prior work.

## 2. Silver Bullet Mechanism

Silver Bullet prevents any RowHammer bit flip by preventively refreshing potential victim rows before they experience any bit flip. Silver Bullet consists of two mechanisms that are responsible for producing (enqueueing) and consuming (performing) preventive refreshes.

### 2.1 Silver Bullet Preventive Refresh Producer Mechanism

Silver Bullet logically divides a DRAM bank into small groups of contiguous rows, called "*subbanks*", which are solely used for bookkeeping purposes. Each subbank is identified by two regions: (1) the preventive refresh region, which contains all the rows in the subbank that must be refreshed by Silver Bullet to avoid RowHammer bit flips, and (2) the counter region, which contains all the rows that, when activated, can cause RowHammer bit flips in the preventive refresh region (i.e., all rows in the preventive refresh region). Silver Bullet maintains a table, called the "*Silver Bullet Table*", that stores an entry for each subbank. Each table entry has three fields: $FRAC$, $PENDING$, and $LOCAL_INDEX$. $FRAC$ and $PENDING$ count the hammers to the subbank's counter region. A subbank's $FRAC$ field is incremented each time a subbank is hammered (i.e., any row in the subbank's counter region is activated, preventively refreshed, or periodically refreshed). When $FRAC$ reaches a threshold value of $D$, *which is a critical parameter of Silver Bullet that needs to be set correctly*, *(1)* the $PENDING$ field is incremented and (2) the $FRAC$ field is reset to 0. $LOCAL_INDEX$ is a field that addresses the next row to refresh. We describe how each table entry is used in RowHammer prevention next.

## 2.2 Silver Bullet Preventive Refresh Consumer Mechanism

Silver Bullet mitigates RowHammer bit flips by effectively increasing the refresh rate (by performing additional *preventive refreshes*) for a subbank at a rate that is correlated with the overall row activation rate in the subbank. The $PENDING$ value indicates the number of preventive refreshes that must be performed on the subbank to prevent RowHammer bit flips. Each preventive refresh targets a single row in the subbank, which is addressed by the $LOCAL_INDEX$ field in the Silver Bullet table entry associated with the subbank. Silver Bullet increments $LOCAL_INDEX$ after performing a preventive refresh on the subbank. This way, Silver Bullet refreshes the subbank's rows one-by-one in a round robin fashion.

Silver Bullet *produces,* or *enqueues,* preventive refreshes for a subbank (i.e., increments the subbank's $PENDING$ value) every time the subbank is hammered $D$ times. This ensures that no row in the subbank can be activated enough times (between two preventive refreshes to the row) to cause a RowHammer bit flip, assuming D is set correctly to account for $UHC_{DRAM}$. Assuming that Silver Bullet is implemented *within* the DRAM chip and assuming *no changes* to the existing DRAM interface, Silver Bullet *consumes,* or *performs,* pending preventive refreshes (i.e., decrements the subbank's $PENDING$ value) at a rate restricted by the DRAM interface protocol (e.g., (LP)DDRX). If there are more than one pending preventive refreshes, Silver Bullet prioritizes the preventive refresh targeting the subbank with the highest $PENDING$ value. To do so, Silver Bullet either (1) scans the Silver Bullet table to find the entry with the highest $PENDING$ value, or (2) maintains a list of subbanks in decreasing $PENDING$ values. In such an implementation, two fields (NEXT and PREV) are added to each entry, such that the order can be maintained (e.g., with a linked list implementation).

## 2.3 Silver Bullet Configuration Parameters

Silver Bullet should be configured by considering five parameters of the target memory subsystem (summarized in Table 1). The target DRAM chip's characteristics define three of these parameters: *unsafe hammer count* ($UHC_{DRAM}$), *blast radius* ($B$) and *bank size* ($S_B$). $UHC_{DRAM}$ is the minimum number of activations that an attacker must perform on a target DRAM row (i.e., aggressor row) to cause a RowHammer bit flip in the rows physically nearby the aggressor row (i.e., victim rows) before the victim rows are refreshed. Blast radius ($B$) is the maximum physical distance (in terms of rows) from the aggressor row at which RowHammer bit flips can be observed. Bank size ($S_B$) is the number of rows in a DRAM bank. Additionally, we define two parameters $R$ and $T$ to model the rate at which Silver Bullet can perform *preventive refreshes, while obeying the constraints of the DRAM interface protocol*. We assume that Silver Bullet can perform preventive refreshes on $R$ rows within every time window that can fit at most $T$ row activations. The values of $T$ and $R$ are constrained by the DRAM protocol, DRAM design, and Silver Bullet implementation.

*Table 1. Target Memory Subsystem Parameters*

| Parameter | Defined by | Definition |
|---|---|---|
| $UHC_{DRAM}$ | Target DRAM Chip | The ***unsafe hammer count*** (i.e., minimum activation count) required to cause a RowHammer bit flip in any row. |
| $B$ | | ***Blast radius:*** The maximum distance (in terms of rows) from the aggressor row at which RowHammer bit flips can be induced. |
| $S_B$ | | **Size of a DRAM Bank:** Number of rows in a DRAM bank |
| $R$ | DRAM Protocol, DRAM Design, Silver Bullet Implementation | Silver Bullet can perform preventive refreshes for $R$ rows within every time window that can fit at most $T$ row activations. For Silver Bullet to be secure, $T$ has to be smaller than $UHC_{DRAM}$. |
| $T$ | | |

Silver Bullet has three key configuration parameters: $D$, $S_{SB}$, and $N_{SB}$ (summarized in Table 2). $D$ indicates the number of hammers that a subbank must receive for Silver Bullet to produce, or enqueue, a preventive refresh to that subbank. $S_{SB}$ indicates the number of rows in each subbank. Therefore, as an approximation[3], a subbank must be hammered $D.S_{SB}$ times before Silver Bullet refreshes the *entire* subbank. $N_{SB}$ indicates the number of subbanks in a bank, which depends on both the bank size and $S_{SB}$. If all subbanks are sized uniformly, $N_{SB} = S_B \,/\, S_{SB}$. In the case that subbanks are *not* equally sized, $N_{SB}$ varies within the range that we specify in Table 2, where $S_{SBmin}$ and $S_{SBmax}$ are the minimum and maximum subbank sizes, respectively.

*Table 2. Key Silver Bullet Configuration Parameters*

| | Constraints | Definition |
|---|---|---|
| $\boldsymbol{D}$ | $D \geq 2(T/R + 1)$ | Number of hammers that a subbank must receive for Silver Bullet to produce, or enqueue, one preventive refresh for the subbank |
| $\boldsymbol{S_{SB}}$ | $2B \leq S_{SB} \leq S_B$ | **Size of a Subbank:** Number of rows in a subbank |
| $\boldsymbol{N_{SB}}$ | $S_B/S_{SBmax} \leq N_{SB}$<br>$N_{SB} \leq 1 + (S_B - S_{SBmax})/S_{SBmin}$ | **Number of Subbanks in a Bank:** in a DRAM bank of size $S_B$, $S_{SBmin}$ and $S_{SBmax}$ are minimum and maximum subbank sizes, respectively. Our analysis, without loss of generality, assumes equal-sized subbanks, i.e., $S_{SBmin} = S_{SBmax}$. |

From the perspective of Silver Bullet, a RowHammer attack sequence is defined at the granularity of subbank hammers (i.e., an activation to any row within a subbank counter region). In this study of Silver Bullet's security guarantees, we mathematically model *any* RowHammer attack on a Silver Bullet-protected system with a minimal number of assumptions in terms of subbank hammer rates and DRAM parameters.

[3] There may be some additional delay between the time that a preventive refresh is enqueued and performed. During this delay, additional hammers can be performed in the subbank. In this proof, we aim to maximize this delay in order to identify the maximum hammer count that an attacker can achieve towards a successful RowHammer attack. We calculate the exact number of hammers in our worst-case attack analysis in Section 6.

# 3. Assumptions and Implementation Constraints

## 3.1 Assumptions in the Proof

In our mathematical analysis and proof, we make the following assumptions, unless otherwise specified.

- We assume that activating an aggressor row affects all rows within its blast radius as strongly as it affects the rows that are physically adjacent to the aggressor row. This assumption defines a pessimistic upper bound for the distribution of RowHammer effect with varying aggressor-victim row distance since prior research clearly shows that the RowHammer effect significantly reduces with the distance between an aggressor and a victim row through conducting experiments on real DRAM chips [3].
- We assume the worst-case scenario in each RowHammer Attack when accounting for refresh operations issued by the standard DRAM refresh protocol [1,4,5] to a subbank. During an attack, we assume that the victim row is never refreshed, but all possible aggressor rows (i.e., $2B$ rows) are refreshed (i.e., activated), increasing the hammer count by $2B$.
- Silver Bullet mitigates RowHammer bit flips in a DRAM chip that uses a standard interface protocol without any modification, e.g., (LP)DDRx. If the DRAM protocol is relaxed and made more flexible (e.g., with DRAM-Initiated Pause), then our proof holds, since a RowHammer attack becomes easier to prevent due to the additional flexibility that allows Silver Bullet to perform any number of preventive refreshes for a given number of activations (i.e., Silver Bullet can be implemented assuming any value of R with arbitrary values of T).
- Silver Bullet is implemented inside the DRAM chip. Our proof also holds for memory controller-based Silver Bullet implementations, because memory controller-based implementations have more relaxed constraints for issuing preventive refresh operations i.e., preventive refreshes can be scheduled very flexibly as activate-precharge pairs by the memory controller (as we discuss in Section 3.2 and evaluate in Section 10.1).
- If there are more than one subbanks with non-zero pending preventive refreshes, Silver Bullet prioritizes performing the preventive refresh on the subbank with the highest $PENDING$ value.
- When performing preventive refreshes on a single subbank, Silver Bullet selects the row to refresh in a round robin fashion. The next row to preventively refresh in a subbank is indicated by the corresponding $LOCAL_INDEX$ field in each Silver Bullet table entry. Each time a preventive refresh is performed on a subbank, the $LOCAL_INDEX$ is incremented while constrained by the subbank size (i.e., $LOCAL_INDEX{=}LOCAL_INDEX + 1 \,\%\, S_{SB}$).
- We examine various prioritization metrics in Silver Bullet for performing preventive refreshes on subbanks that have the same *maximum* $PENDING$ value (i.e., the highest $PENDING$ value across all subbanks). We show the proof for the worst-case prioritization scheme to demonstrate the security guarantees of Silver Bullet in all possible implementations. In the worst-case prioritization scheme, Silver Bullet always prioritizes a non-target subbank over the target subbank (i.e., Silver Bullet must perform preventive refreshes on all non-target subbanks with the same $PENDING$ value as the target subbank before refreshing the target subbank).

- Our main proof assumes that subbanks do not correspond to physically compartmentalized groupings of DRAM rows (i.e., activating a row in one subbank can potentially cause RowHammer bit flips in another subbank). To account for cross-subbank RowHammer attacks, Silver Bullet suggests two alternative schemes: 1) extended counter region scheme and 2) extended preventive refresh region scheme.

  **Extended Counter Region Scheme.** The first scheme increments a subbank's $FRAC$ value upon a row activation, if the activated row is within the blast radius of any of the rows in the subbank. To do so, this scheme defines the counter region of a subbank such that the counter region contains both (i) the rows within the subbank and (ii) the rows outside of the subbank but within the blast radius of any rows within the subbank.

  **Extended Preventive Refresh Region Scheme.** The second scheme increments a subbank's $FRAC$ value upon a row activation only if the activated row is within the subbank boundaries. In this scheme the subbank's preventive refresh region is extended such that Silver Bullet performs additional preventive refreshes on all rows within the blast radius of rows within the subbank. In our main proof, we consider the first scheme (i.e., extended counter region) since it is more desirable for realistic subbank sizes.[4] Note that when a subbank's $FRAC$ counter is incremented as a result of performing an activation, a preventive refresh, or a periodic refresh on a row within the subbank, we consider the subbank to have been hammered (i.e., subbank hammer). Therefore, when a single activation is performed in an extended counter region, multiple subbanks have been hammered. Figure 1 depicts the extended counting region scheme.

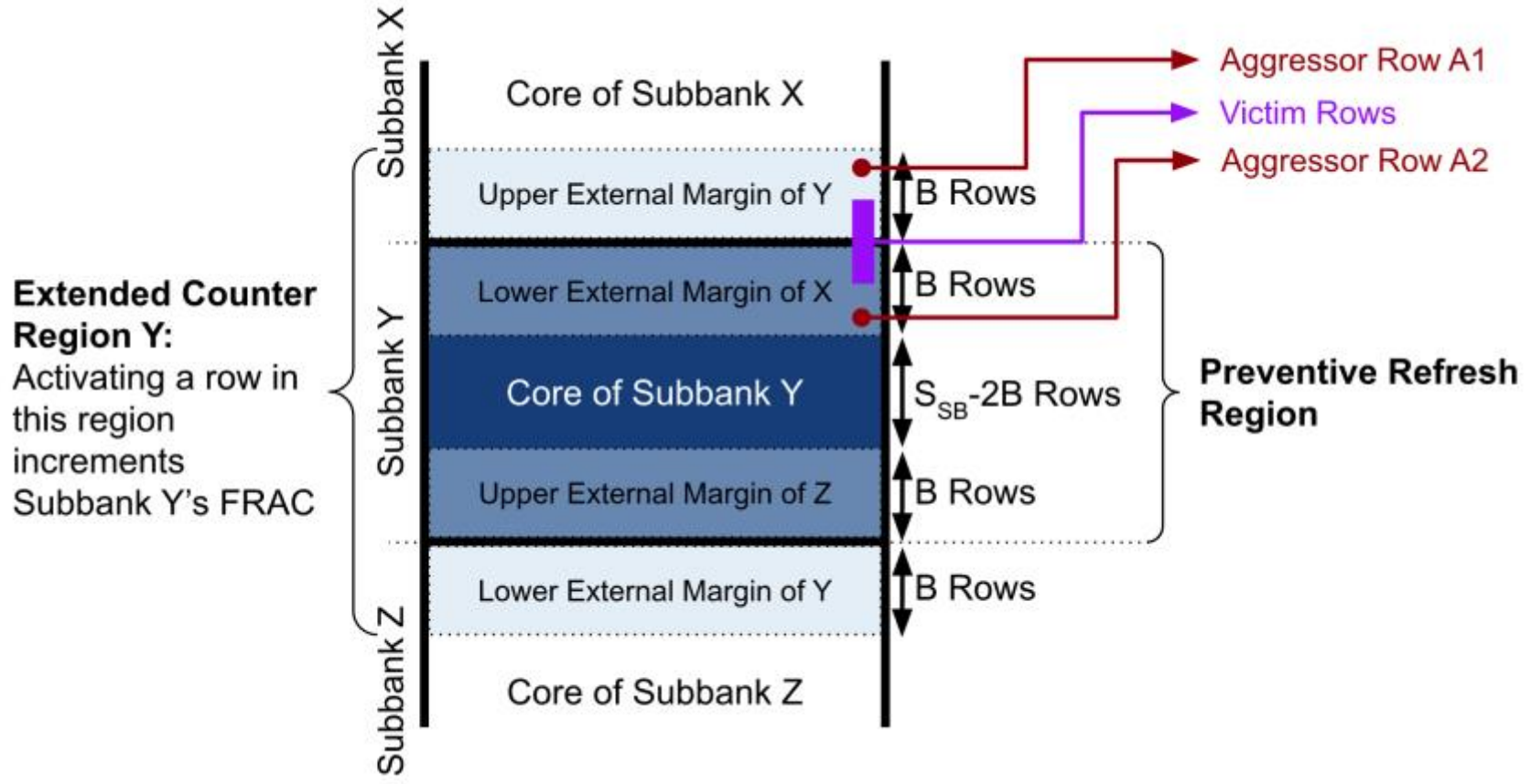

*Figure 1. Extended Counter Region Subbank Overlapping Scheme*

Figure 1 shows three subbanks (i.e., Subbanks X, Y, and Z, drawn with bold lines) that are physically adjacent to each other, organized according to the extended counter region scheme. In this scheme, Subbank Y's FRAC counter is incremented when a row in

[4] We study the second scheme (i.e., extended preventive refresh region) in Appendix B.

Subbank Y's extended counter region is activated (i.e., Subbank Y and its upper and lower external margins in Figure 1). Subbank Y is completely refreshed when Silver Bullet performs a preventive refresh on each row in Subbank Y's preventive refresh region. In this scheme, an activation to aggressor row A1 or A2 increments the $FRAC$ counters of both Subbanks X and Y. This enables Silver Bullet to perform preventive refreshes on any potential victim row nearby aggressor rows A1 and A2 in both Subbanks X and Y.

## 3.2 Consumption Rate of Pending Preventive Refreshes

We model the rate at which Silver Bullet can perform preventive refreshes with parameters $R$ and $T$. We assume that Silver Bullet can perform preventive refreshes on $R$ rows after every time window that can fit at most $T$ row activations. Parameters $R$ and $T$ are set in one of the following four ways based on the memory subsystem design.

First, $R$ and $T$ can be set according to the refresh latency $(t_{RFC})$ and refresh interval $(t_{REFI})$, defined by the DRAM standard (e.g., (LP)DDRX). The memory controller already allocates a $t_{RFC}$ time window for regular refreshes after every $t_{REFI}$ time window during which it services memory requests. By leveraging (the time slack, refresh-level internal parallelism, or both in) $t_{RFC}$, Silver Bullet can consume $R$ preventive refreshes after every $t_{REFI}$ time window. Although the $t_{RFC}$ time window is dedicated for refresh operations and memory requests are serviced only between refresh commands (i.e., within $t_{REFI}$), we extend these constraints in favor of the attacker to build our proof based on a theoretical worst case. In this worst case, we assume that the attacker can perform activations during both $t_{REFI}$ and $t_{RFC}$ time windows. Therefore, we define $T$ as the maximum number of activations that can be issued in a $t_{REFI} + t_{RFC}$ time window. Note that the DRAM interface protocol defines $t_{RFC}$ and $t_{REFI}$, but the slack and refresh-level parallelism in $t_{RFC}$ is dependent on the DRAM design and implementation. It is possible to perform multiple refresh operations in parallel in a single DRAM bank by leveraging the parallelism across physically-separated groupings of DRAM rows, which are referred to as *subarrays* in the literature. Each subarray contains its own intermediary sense amplifiers (i.e., local sense amplifiers), allowing multiple subarrays to simultaneously perform refresh operations without interfering with each other. In a bank that consists of N subarrays, N rows can be simultaneously refreshed (one per subarray), assuming a sufficient power budget. We provide a detailed example of this parallelism in Appendix C.

Second, $R$ can be set based on reserved periods of time that two recent DRAM standards provide for RowHammer mitigation operations: HBM's TRR mode [4] and DDR5's RFM (i.e., refresh management) command [5]. These implementations enable the memory controller to dedicate a certain time window for an on-die RowHammer mitigation mechanism to perform preventive refreshes. In this design, the memory controller and the DRAM chip design define parameter $R$.

Third, Silver Bullet can leverage the DRAM initiated pause (DIP) feature, proposed by UPMEM. This feature enables the DRAM chip/module to communicate with the memory controller, asking it to pause issuing commands to the DRAM chip/module for a certain amount of time. Silver Bullet can use this feature to prevent RowHammer bit flips by pausing the memory controller at arbitrary points in time and performing pending preventive refreshes as necessary. When using DIP, Silver

Bullet can (during online operation) set the value of $R$, independently of the tight restrictions imposed by the standard DRAM protocols.

Fourth, to avoid the limitations on $R$ and $T$ imposed by the DRAM interface protocol, Silver Bullet can be implemented in the memory controller such that Silver Bullet can schedule activation-precharge pairs to rows for which a preventive refresh is required. In this implementation, Silver Bullet's $R$ and $T$ parameters can be set freely.

Clearly, the first scenario is the most stringent in terms of requirements imposed on Silver Bullet. If Silver Bullet can guarantee secure operation under the first scenario, then it can much more easily guarantee secure operation under the other three scenarios. As such, our proof assumes, without loss of generality, the first scenario.

## 3.3 Production and Consumption Rate Constraints on Preventive Refreshes

Silver Bullet *produces,* or *enqueues,* one preventive refresh (i.e., increments the $PENDING$ value) for a subbank for every $D$ subbank hammers. Because Silver Bullet extends the counter region of each subbank, the counter regions of adjacent subbanks overlap. Therefore, hammering (i.e., activating or preventively refreshing) a row D times can produce one preventive refresh for each of *two* adjacent subbanks if the row is in the overlapping part of the adjacent subbanks' extended counter regions. Silver Bullet *consumes,* or *performs,* one preventive refresh (i.e., decrements the $PENDING$ value) at a rate dictated by $R/T$. To prevent overflowing the $PENDING$ counter (and thus to operate correctly), a Silver Bullet implementation must be configured such that the rate at which preventive refreshes are produced is *not* larger than the rate at which preventive refreshes are consumed. Within a time window of T activations: (1) the attacker's $T$activations can produce $2T/D$preventive refreshes, (2) Silver Bullet can consume $R$ preventive refreshes, and (3) the $R$ preventive refreshes performed by Silver Bullet produce $2R/D$ additional preventive refreshes. This leads us to the first constraint, shown in Expression 1, that a Silver Bullet implementation must satisfy for correct operation.

$$Produced\ preventive\ refresh\ count\ \leq Consumed\ preventive\ refresh\ count$$

$$2T/D\ +\ 2R/D\ \leq R$$

We solve this inequality for $D$:

$$D \geq 2(T/R\ + 1) \ldots\ldots\ldots\ldots\ldots\ldots\ldots\ldots\ldots\ldots\ldots\ldots\ldots\ldots (Expression\ 1)$$

**Observation 1.** Silver Bullet must be configured with respect to the lower bound for the parameter $D$, defined in Expression 1 as $D \geq 2(T/R\ + 1)$.

## 4. RowHammer Attack Model on a Silver Bullet-Protected System

Our goal is to prove that a Silver Bullet-protected system is secure (i.e., RowHammer-safe, that is, free of RowHammer-induced bitflips) against the strongest possible attacker model. We consider the following strongest attacker model against a Silver Bullet-protected system and describe the constraints under which an attacker can operate, before we build our worst-case attack in Section 5.

## 4.1 Attacker Model

There exist several prior studies on the effectiveness of various DRAM access patterns in causing RowHammer bit flips. These access patterns include 1) *single-sided attacks*, where a single row that is adjacent to the victim row is repeatedly activated, and 2) *double-sided attacks*, where the two physically adjacent rows to the victim row are repeatedly activated in an alternating fashion. Our attacker model includes all possible access patterns to ensure that Silver Bullet is secure against *all* possible RowHammer attacks. This includes the worst-case access pattern, which causes RowHammer bit flips in the DRAM chip using the lowest possible number of activations (i.e., $UHC_{DRAM}$) performed on the worst-case set of aggressor rows with the worst-case access pattern. We refer to the subbank containing the victim row as the *target subbank*.

## 4.2 Modeling Silver Bullet at the Subbank Granularity

We model all possible access patterns that a RowHammer attack can exhibit from the perspective of $FRAC$ and $PENDING$ values by leveraging two features of Silver Bullet. First, for any victim row in a subbank, all rows that can contribute to a RowHammer bit flip in the victim row (i.e., rows within the blast radius) increment the FRAC counter associated with the victim subbank when activated (or refreshed). Preventive refreshes are subsequently performed according to each subbank's activation count (i.e., $FRAC$ and $PENDING$). Second, Silver Bullet performs preventive refreshes to the subbank proportionally to the rate at which the subbank's hammer counter is incremented. This ensures that Silver Bullet refreshes *all rows* in a subbank *before* the total hammer count of the subbank reaches a specified value, assuming Silver Bullet has enough time to perform all the required preventive refreshes.

These features enable us to validate the security guarantees of Silver Bullet by ensuring that no single subbank can reach a hammer count higher than $UHC_{DRAM}$ between two subsequent preventive refresh operations to the victim row in the subbank. If an attacker can achieve a hammer count higher than $UHC_{DRAM}$ in a subbank, the attacker can perform the worst-case RowHammer attack and induce RowHammer bit flips. Otherwise, the system is RowHammer-safe.

## 4.3 Maximizing the Silver Bullet Preventive Refresh Window

The *preventive refresh window* (i.e., the time window between two subsequent preventive refreshes to a single row) of a subbank limits the maximum subbank hammer count. The preventive refresh window depends on 1) the number of rows in the subbank, 2) the $PENDING$ values of *other* subbanks, and 3) the blast radius that the DRAM chip exhibits. We next discuss these two Silver Bullet features that an attacker should account for when maximizing the subbank hammer count during a preventive refresh window.

Since Silver Bullet performs preventive refreshes to rows in a subbank in round robin order, an attacker should select the row that was most recently refreshed by Silver Bullet as the victim row (i.e., the row to induce RowHammer bit flips in) just before starting the RowHammer attack sequence. This ensures that the attacker can maximize the target subbank's hammer count before the victim row is refreshed by Silver Bullet (i.e., within a preventive refresh window). This maximum hammer count is calculated by multiplying the number of rows in a subbank $(S_{SB})$ by the number of activations $(D)$ required for Silver Bullet to perform a preventive refresh for a single row in that subbank: $D.S_{SB}$

**Opportunity 1.** An attacker should select the row that was most recently refreshed by Silver Bullet, as the victim row, to maximize the hammer count during any single preventive refresh window.

Second, Silver Bullet prioritizes preventive refreshes to the subbank with the highest $PENDING$ value in a bank. Therefore, an attacker can extend the preventive refresh window that a target subbank experiences by using an access pattern that causes Silver Bullet to prioritize preventive refreshes to *non-target subbanks*, i.e., subbanks that are not the target subbank. This is only possible by increasing the $PENDING$ value of non-target subbanks. Next, we explain how the number of non-target subbanks with a given $PENDING$ value affects the preventive refresh window that the target subbank experiences.

Consider the case where $n$ subbanks have the highest $PENDING$ value in a bank. When Silver Bullet performs the first preventive refresh in one of those subbanks (say subbank A), subbank A's $PENDING$ value is decremented by one and the number of subbanks with the highest $PENDING$ value becomes $n-1$. Silver Bullet performs the next preventive refresh in one of the remaining $n-1$subbanks. With this scheme, Silver Bullet can only perform a second preventive refresh on subbank A after issuing one preventive refresh to *every other* subbank with the highest $PENDING$ value (i.e., $n-1$ different subbanks). In this example, the preventive refresh window of subbank A is extended due to the $PENDING$ values of the other subbanks. An attacker can exploit this subbank prioritization feature of Silver Bullet to effectively reduce the rate at which preventive refreshes are performed in a target subbank and thus effectively extend the preventive refresh window. To maximize the preventive refresh window, an attacker should aim to maximize the $PENDING$ values of non-target subbanks. This reduces the rate at which preventive refreshes are performed in the target subbank.

**Opportunity 2.** An attacker can extend the preventive refresh window that a target subbank experiences by increasing the $PENDING$ value of other subbanks.

Since the number of rows in a subbank is fixed by the Silver Bullet implementation, in the attack formulation we construct (in Section 5), we explore methods for increasing the $PENDING$ value across as many subbanks as possible to maximize the preventive refresh window that the target subbank experiences.

# 5. Generalized RowHammer Attack Formulation

We construct the Wave Attack, a generalized attack that aims to maximize the hammer count of a subbank. Without loss of generality, an attacker can maximize the subbank hammer count by extending the target subbank's preventive refresh window (see Opportunity 2 and its preceding explanation in Section 4.3). We construct a general worst-case RowHammer attack in two phases:

**Phase 1.** Maximize the $PENDING$ value $(P_{max})$ across a set of subbanks to maximize the preventive refresh window that the target subbank experiences.

**Phase 2.** Maximize the hammer count of a target subbank within a preventive refresh window (i.e., before Silver Bullet refreshes every row in the target subbank).

To construct the worst-case attack pattern, we individually construct the worst-case for each phase. First, we create the conditions for and calculate the worst-case $P_{max}$that Phase 1 can achieve. According to our analysis in Section 5.1, an attacker can increase the $PENDING$ value of only a subset of subbanks to $P_{max}$. Second, in Section 5.2, we create the conditions for and calculate the worst-case hammer count that an attacker can reach in the target subbank within a preventive refresh window. We conduct this analysis with a conservative assumption in which an infinite number of subbanks have the $PENDING$ value of $P_{max}$. This lets us identify the theoretically highest possible hammer count of a subbank within a preventive refresh window. Silver Bullet can only provably-securely prevent RowHammer bit flips on DRAM chips that have a $UHC_{DRAM}$ value greater than the identified highest possible subbank hammer count.

## 5.1 Phase 1: Maximizing the Highest PENDING Value Across Subbanks

In **Phase 1**, the attacker's primary goal is to maximize the highest $PENDING$ value across as many subbanks as possible within the bank that contains the target subbank.

Effectively increasing the highest $PENDING$ value $(P)$ in a bank is not possible by hammering only a single subbank due to the constraint $D \geq 2(T/R+1)$ in *Expression 1*. When hammering a single subbank, the attacker can increase the subbank's $PENDING$ value to up to $(T+R)/D$ during a given time window of $T$activations. At the end of this time window, Silver Bullet is capable of performing up to $R$ consecutive preventive refreshes, decrementing the $PENDING$ value by $R$. For Silver Bullet to consume all preventive refreshes that are produced within T activation time window, $R$ should be larger than or equal to $(T+R)/D$. Therefore, we evaluate the conditions where the inequality of $R \geq (T+R)/D$is satisfied. By reorganizing the inequality, we derive $R(D-1) \geq T$. When D takes the minimum value it can take based on Expression 1 (i.e., $D_{min} = 2(T/R+1)$), the inequality evaluates $2 \geq 1$ and left hand side of the inequality grows as $D$ takes larger values. Therefore, the inequality $R \geq (T+R)/D$ is satisfied for all sets of parameters that satisfy Expression 1. Thus, we conclude that the $PENDING$ value resets to zero after every $T$ activations, and it can never exceed $R$ during this time window, thereby limiting $P$ to $(T+R)/D$ , when only a single subbank is hammered.

To effectively increase the highest $PENDING$ value in a bank, the attacker should maximize the number of subbanks whose $PENDING$ values are incremented to the new highest $PENDING$ value *(i.e., $P+1$)* before Silver Bullet performs preventive refreshes on them. Since Silver Bullet performs (consumes) preventive refreshes at a rate equal to or greater than they are created (produced) (*Expression 1*), *the attacker must minimize the time spent for increasing the* $PENDING$ values *across all subbanks* (as explained in Section 5.1.2). Note that the attacker can increase the $PENDING$ value of two adjacent subbanks from $P$ to $P+1$by hammering any of the rows contained in both of the subbanks' extended counter regions $D$ times.

### 5.1.1 Modeling All Possible Access Patterns for Phase 1

In order to maximize the $PENDING$ value for any subbank, we develop a set of steps (as shown in Figure 2 and described below), that attempt to overwhelm Silver Bullet (i.e., cause it to generate more preventive refreshes than can be consumed) by exploiting two observations: (1) Silver Bullet only performs preventive refresh operations when there are subbanks with non-zero $PENDING$ values and (2) Silver Bullet always prioritizes a subbank with the highest $PENDING$ value across the bank when performing preventive refreshes. In this section, we model this phase of the attack (and any variation of it), which aims to maximize the highest $PENDING$ value $P$and subsequently the subbank hammer count of the target subbank within a preventive refresh window.

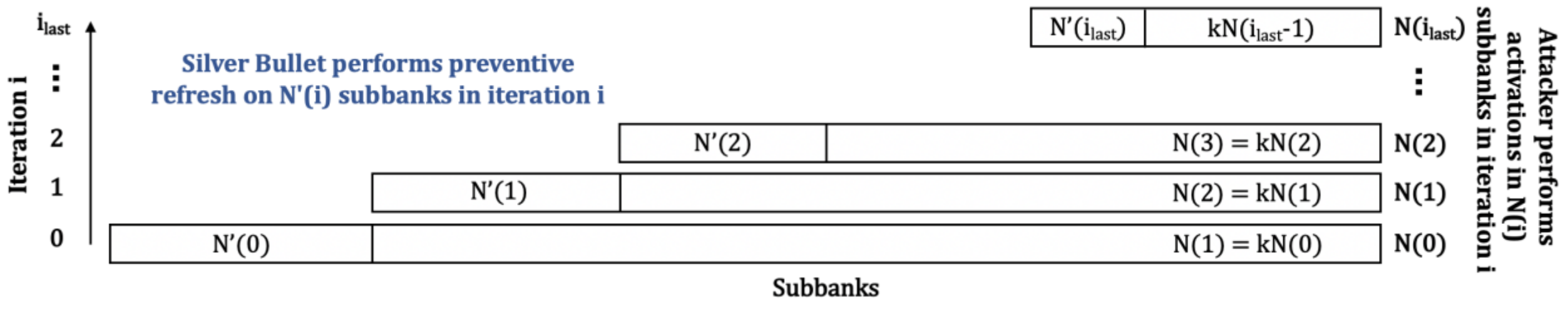

*Figure 2. Phase 1 of the attack. In each iteration, the number of subbanks with the highest $PENDING$ value ($N(i)$) decreases, since Silver Bullet refreshes $N'(i)$ subbanks. The rate of reduction in subbanks from one iteration to the next is $k$.*

**Initial state:** An aggressor row in every subbank is already activated $D-1$ times, which is the maximum number of activations that can be performed without triggering any preventive refresh by Silver Bullet.

**In each iteration *i* of Phase 1, the attacker does the following:**

1. The attacker identifies a number of subbanks to attack in iteration $i$ (i.e., $N(i)$).
2. The hammer count of each of these $N(i)$ subbanks is incremented by 1 due to an activation or a refresh, which is performed on each of the $N(i)/2$ overlapping extended counter regions. Doing so consequently produces a new pending preventive refresh (i.e., $PENDING++$) and clears the subbank activation counter ($FRAC=0$) for each of the $N(i)$subbanks.
3. The attacker identifies the set of subbanks to attack in the next iteration. We represent the number of subbanks that will be hammered in the next iteration as $N(i+1)$. We do not define any constraints at this point for the sets of subbanks in iterations $i$ and $i+1$.

4. The hammer count of each of these $N(i+1)$subbanks is incremented by $D-1$due to row activations and refreshes. To do so, the attacker performs just enough activations on each of the $N(i+1)/2$ overlapping extended counter regions such that the subbank's hammer count reaches $D-1$. Let $N_{ACT}(i)$ be the sum of the (1) total number of activations that the attacker has to issue and (2) total number of preventive refreshes performed by Silver Bullet in iteration $i$. We calculate $N_{ACT}(i)$ in *Expression 2* as the sum of the activation counts from *steps 2 and 4*.

$$N_{ACT}(i) = N(i)/2 + (D-1)N(i+1)/2 \ldots\ldots (Expression\ 2)$$

**During each iteration of Phase 1:**

- Let $N'(i)$ be the number of preventive refreshes performed in iteration $i$. Given that Silver Bullet performs $R$ preventive refreshes for every $T$activations (the ones issued by the attacker and the ones issued by Silver Bullet itself), we derive $N'(i)$, as follows, in *Expression 3*:

$$N'(i) = (R/T)N_{ACT}(i), where\ N_{ACT}(i) = N(i)/2 + (D-1)N(i+1)/2\ \ from\ Expression\ 2$$

$$N'(i) = \frac{R}{2T}(N(i) + (D-1)N(i+1)) \ldots\ldots (Expression\ 3)$$

### 5.1.2. Intelligently Selecting the Subbanks in Every Iteration

During Phase 1, the goal of the attacker is to maximize the highest $PENDING$ value. Therefore, in each iteration, the attacker should aim to increase the current highest $PENDING$ value, $P$, to $P+1$, and maximize the number of subbanks with the new highest $PENDING$ value of $P+1$. The attacker should consider three Silver Bullet features when attempting to achieve this goal.

First, after initialization (i.e., increasing the $FRAC$ value of every subbank to $D-1$), Silver Bullet performs preventive refreshes (decrements $PENDING$ values) at a rate equal to or higher than the rate an attacker can produce (i.e., increment $PENDING$ values) them (from *Expression 1*). Due to this discrepancy in production/consumption of preventive refreshes, the attacker has limited time to maximize the highest $PENDING$ value in subbanks across the bank before all pending preventive refreshes are consumed (i.e., the total $PENDING$ value across the bank is zero). Therefore, the attacker should intelligently allocate its limited number of activations (before all subbanks with a non-zero $PENDING$ value are refreshed by Silver Bullet) to select subbanks in every iteration to achieve the goal of Phase 1.

Second, the required number of activations to increase the pending value of a subbank to $P+1$ is proportional to the difference between $P+1$ and the current $PENDING$ value of the subbank. To minimize the time spent for increasing the $PENDING$ value during each iteration, the attacker should only hammer subbanks that already have the current highest $PENDING$ value (i.e., $P$), such that performing only $D/2$ activations or preventive refreshes (since one activation hammers or increases the counters of two subbanks) per subbank are required to increase their $PENDING$ values to $P+1$.

Third, Silver Bullet prioritizes performing preventive refreshes on the $N(i)$ subbanks with the current highest $PENDING$ value ($P$) across the bank. Therefore, the attacker should issue D activations only to a subset of subbanks ($N(i+1)$) with the current highest $PENDING$ value such that the attacker can increment the $PENDING$ value of each of the chosen subbanks (to $P+1$) while Silver Bullet is performing preventive refreshes on the remaining $N'(i)$ subbanks with the current highest $PENDING$ value ($P$).

If the attacker follows this selection criteria of subbanks to activate in each iteration i, $N(i) - N'(i)$ subbanks should have the new highest $PENDING$ value ($P+1$) at the end of each iteration $i$. In the subsequent iteration, the attacker should choose a set of subbanks from these $N(i) - N'(i)$ subbanks to hammer in iteration $i+1$. Therefore:

$$N(i+1) \leq N(i) - N'(i) \ldots\ldots (Expression\ 4)$$

We substitute $N'(i)$ in *Expression 4* with its definition in *Expression 3* and solve for $N(i+1)$ to determine the number of subbanks whose $PENDING$ values are incremented in iteration $i$, as follows:

$$N(i+1) \leq N(i) - (R/2T)\ (N(i) + (D-1)N(i+1))$$

$$N(i+1)(1 + (D-1)(R/2T)) \leq (1 - (R/2T))N(i)$$

$$We\ can\ make\ the\ following\ simplification\ since\ both\ D > 1\ and\ T > 1$$

$$N(i+1) \leq k\ N(i), where\ k = \frac{2T - R}{2T + (D-1)R} \ldots\ldots (Expression\ 5)$$

*Expression 5* defines a parameter called *k. k* is the reduction factor in the number of subbanks from one iteration ($i$) to the next ($i+1$). It determines the number of iterations that Phase 1 can last before there are no subbanks left. We analyze the possible values of $k$ to calculate the maximum $PENDING$ value a subbank can reach.

### 5.1.3. Analysis of the Subbank Count Reduction Factor (k)

We calculate the range of values that the subbank count reduction factor ($k$) can take.

$$k = \frac{2T - R}{2T + (D-1)\ R} = \frac{2 - R/T}{2 + (D-1)R/T}\ from\ Expression\ 5$$

We solve this equation for $R/T$ :

$$R/T = 2(1-k)/(kD - k + 1)$$

Given that $2/D \leq R/T$(Observation 1):

$$2/D \leq 2(1-k)/(kD - k + 1)$$

We solve this inequality for $k$:

$$k \leq (D-1)/(2D-1) \ldots\ldots (Expression\ 6)$$

To further understand the subbank count reduction at each iteration, we calculate the upper bound of $k$ for various values of parameter $D$.

By definition, D is a positive integer. Since upper bound of $k$ increases with $D$ based on Expression 6, we calculate a theoretical upper bound for $k$ by taking the limit of Expression 6 as D approaches infinity.

$$lim_{D\to\infty}(D-1)/(2D-1) = 0.5$$

$$0 < k \quad < 0.5 \ldots\ldots (Expression\ 7)$$

From *Expression 7*, an attacker should increment the $PENDING$ value of at most half of the subbanks with the current highest $PENDING$ value in each iteration in order to maximize the highest $PENDING$ value in Phase 1.

Therefore, we define an upper bound constraint that the attacker should obey to maximize the highest $PENDING$ value in Phase 1 with the goal of overwhelming Silver Bullet's prevention mechanism. To do so, we solve *Expression 5* for $N(i)$ by using Expression 7 and derive *Expression 8*.[5]

$$N(i) \le k^i N(0), where\ i > 0\ and\ 0 < k < 0.5 \ldots\ldots (Expression\ 8)$$

**Observation 2.** The attacker must reduce the number of subbanks it hammers in every iteration, with respect to *Expression 8*.

#### 5.1.4 Calculating the Maximum Possible $PENDING$ Value ($P_{p1max}$)

The first phase of the attack can be sustained for a limited number of iterations due to the monotonically decreasing number of subbanks in each iteration (see Observation 2). We calculate the maximum number of hammers that an attacker can perform on the target subbank in the first phase based on Observation 2. Let $i_{last}$ be the last iteration of the first phase, defined by two properties: (1) during this iteration, the attacker can still hammer at least one subbank, but (2) after this iteration (i.e., in a hypothetical iteration $i_{last+1}$), the attacker cannot hammer any subbank due to the monotonically decreasing number of subbanks in each iteration.

The attacker should be able to hammer at least one subbank in iteration $i_{last}$: $N(i_{last}) \ge 1 and\ \ N(i_{last}) \le k^{i_{last}} N(0)\ from\ Expression\ 8$

$$k^{i_{last}} N(0) \ge 1$$

Solving for $i_{last}$:

$i_{last} . log(k) \ \ge\ log(1/N(0))$ Note that $k < 0.5$from *Expression 7*. Thus, $log(k)\ <\ 0$

$$i_{last} \le log(1/N(0))\ /\ log(k)$$

$$i_{last} \le log_k(1/N(0)) \ldots\ldots (Expression\ 9)$$

---

[5] Note that *k* is constant across all iterations because it is defined by the values of *R*, *T*, and *D* as stated in *Expression 5*.

According to *Expression 9*, Phase 1 can not last longer than $log_k(1/N(0))$iterations. Note that the highest $PENDING$ value in the bank is incremented by one every iteration. Therefore, the maximum $PENDING$ value that an attacker can reach in Phase 1, i.e., $P_{p1max}$, is $log_k(1/N(0))$.

$$P_{p1max} = log_k(1/N(0))\ldots\ldots\ldots\ldots\ldots\ldots\ldots\ldots\ldots\ldots\ldots\ldots(Expression\ 10)$$

**Observation 3.** An attacker can perform the first phase for at most $i_{last}$iterations, as specified by *Expression 9*. In the worst-case, the maximum $PENDING$ value in the bank can reach $log_k(1/N(0))$ based on *Expression 10*.

Next, we calculate the worst-case configuration of the access pattern in Phase 1 to achieve the highest possible $PENDING$ value. To do so, we investigate how changing $k$ and $N(0)$ affects $P_{p1max}$. Intuitively, increasing $k$ or $N(0)$ increases the number of subbanks that an attacker can hammer at any given iteration based on *Expression 8*, which consequently increases the number of iterations that Phase 1 can last and the maximum $PENDING$ value that can be reached in Phase 1.

*Expression 10* agrees with this intuitive expectation. Since both $k$ and $1/N(0)$ are strictly smaller than 1, $P_{p1max}$ increases with larger k and N(0) values based on *Expression 10*. We visualize the the relationship between $D, k, N(0)$, and $P_{p1max}$in Figure 3.

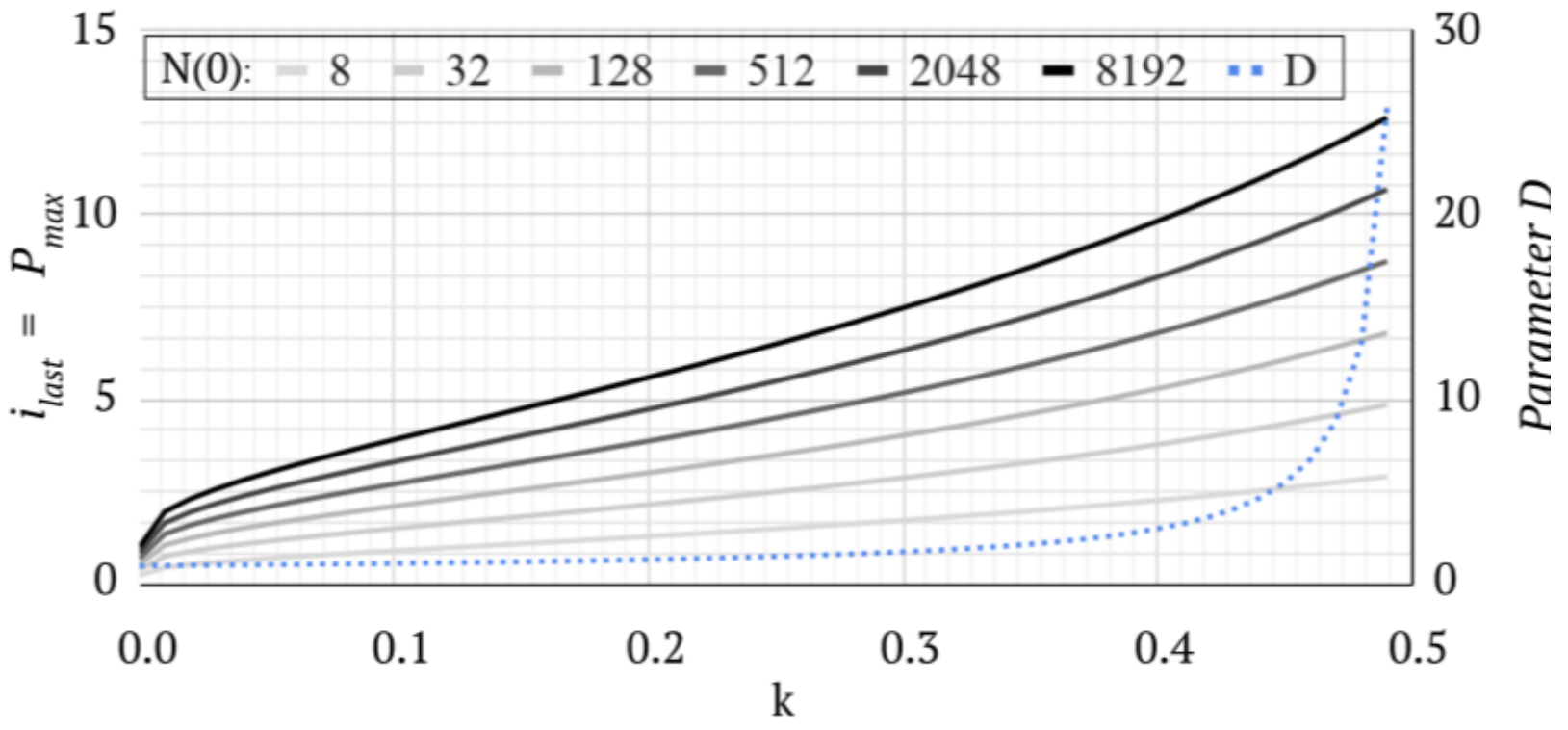

*Figure 3. Maximizing iteration count (and thus the maximum $PENDING$ value) in Phase 1 of the attack.*

Figure 3 shows that increasing $k$ or $N(0)$ monotonically increases $P_{p1max}$. Therefore, a larger $k$ and larger $N(0)$ are better from the perspective of the attacker.

$k$ has an upper bound ($k_{max}$) defined by parameter $D$ (*Expression 6*). Figure 3 shows that for realistic values of $D$ ($D \geq 2$), $k_{max}$ is in the range between 0.33 and 0.5. *Expression 7* shows that $k_{max}$never reaches 0.5 for any practical value of $D$, i.e., the line $x = 0.5$is an asymptote for the $y = D$curve, as seen in Figure 3. Based on this observation, to make sure our analysis covers the worst case, we safely assume $k = k_{max} = 0.5$.

$N(0)$ is the number of subbanks that the attacker chooses to initialize in Phase 1. In the worst case, the attacker initializes all subbanks in the bank, and thus sets $N(0)$ to be the number of subbanks in the bank, which we denote as $N_{SB}$. Selecting all available subbanks maximizes the initial number of pending preventive refreshes (in iteration 0). This is because immediately following initialization, we can produce two preventive refreshes per activation (for the first $N_{SB}$ activations), by activating every subbank once (i.e., incrementing the $FRAC$ value of each subbank from $D-1$ to $D$).

We calculate a theoretical upper bound for $P_{max}$ by applying these worst-case estimations for $k$ and $N(0)$ on *Expression 10*, thereby deriving *Expression 11*, as follows:

$$P_{p1max} = log_k(1/N(0)) \quad from\ Expression\ 10$$

$$k = k_{max} = 0.5\ and\ N(0) = N_{SB}\ (theoretical\ upper\ bounds\ for\ k\ and\ N(0)\ from\ Figure\ 3)$$

$$P_{p1max} = log_{0.5}(1/N_{SB})$$

$$P_{p1max} = (log1 - logN_{SB})\ /\ (log1 - log2)$$

$$P_{p1max} = log_2(N_{SB}) \ldots\ldots\ldots\ldots (Expression\ 11)$$

*Expression 11* suggests that the maximum $PENDING$ value that an attacker can reach in Phase 1 scales logarithmically with the number of subbanks in a bank. This logarithmic relationship is because the number of subbanks with the current highest $PENDING$ value is halved in each iteration after initialization.

According to *Expression 11*, to mount the worst-case attack, an attacker should halve the number of subbanks it hammers after every iteration. As we describe now, this is also intuitive given that the preventive refreshes are produced and consumed at the same rate (i.e., the worst case according to *Expression 1*: $D = 2(T/R + 1)$), which is the minimum requirement for correct operation of Silver Bullet. Assuming preventive refreshes are produced and consumed at the same rate, in a given iteration, Silver Bullet performs preventive refreshes to exactly the same number of subbanks as the number of subbanks whose $PENDING$ value is incremented by the attacker. As such, the attacker increases the $PENDING$ values in half of the subbanks with the current highest $PENDING$ value in each iteration, while at the same time, Silver Bullet decreases the $PENDING$ value of the other half of the subbanks by performing a preventive refresh on each subbank. Thus, we draw our Observation 4 based on *Expression 11* (and its intuitive explanation we just provided):

**Observation 4.** The theoretical upper bound for the highest $PENDING$ value that an attacker can achieve on a Silver Bullet-protected system scales logarithmically with the number of subbanks.

By the end of Phase 1 of the Wave attack, the attacker has maximized the highest $PENDING$ value across the bank to maximize the preventive refresh window for the target subbank, thereby maximizing the number of activations that can be performed in the target subbank before Silver Bullet performs preventive refreshes in the entire subbank. Phase 2 begins when the $PENDING$ value reaches its maximum value (i.e., $P_{p1max}$).

**Hammer Count from Phase 1 (*HC₁*).** An attacker performs $D$ hammers on the target subbank in each iteration. Say, the target subbank experiences a preventive refresh when its $PENDING$ value is $P_{REF}$. In our attack model, we do not make any assumption about *how* the attacker prioritizes hammering the target subbank in Phase 1. Therefore, we consider that $P_{REF}$ can be any integer value up to $P_{max}$.

$$1 \leq P_{REF} \leq P_{p1max} \ldots\ldots (Expression\ 12)$$

At the end of Phase 1, the target subbank is in either of the following two states.

First, the target subbank receives a preventive refresh during Phase 1 when its PENDING value is $P_{REF}$. Until receiving the preventive refresh, the target subbank is hammered for $D.P_{REF}$ times. When Silver Bullet refreshes the target subbank, its $PENDING$ value is decremented to $P_{REF} - 1$. The target subbank does not receive a second preventive refresh in Phase 1 because there is always a subbank with a larger $PENDING$ value on which Silver Bullet performs preventive refreshes. Therefore, at the end of Phase 1, the target subbank's hammer count $(HC_1)$ reaches $D.P_{REF}$, the target subbank's $PENDING$ value becomes $P_{REF} - 1$, and completely refreshing the target subbank requires performing $S_{SB} - 1$ preventive refreshes. *Expression 13* shows the $HC_1$ calculation for this case.

$$HC_1 = D.P_{REF}, \quad if\ 1 \leq P_{REF} \leq P_{p1max} \ldots\ldots (Expression\ 13)$$

Second, the target subbank does *not* receive a preventive refresh in Phase 1, and thus reaches the maximum $PENDING$ value $P_{p1max}$ .In this case, the target subbank's hammer count $(HC_1)$ reaches $D.\ P_{p1max}$ and completely refreshing the target subbank requires performing $S_{SB}$ preventive refreshes. *Expression 14* shows the $HC_1$ calculation for this case.

$$HC_1 = D\ .P_{p1max} \ldots\ldots (Expression\ 14)$$

We demonstrate in Appendix A, that with our worst-case assumptions, the $P_{REF}$ value of the selected target subbank does *not* affect the total hammer count achievable by an attacker. Therefore, we continue our proof assuming that Silver Bullet does *not* perform any preventive refreshes to the target subbank during Phase 1 for simplicity. Thus, we use Expression 14 to calculate $HC_1$ in the proof in the main document text below. Appendix A demonstrates our proof for the alternative case where Silver Bullet does perform a preventive refresh to the target subbank in Phase 1. For this alternative case, we use Expression 13 to calculate $HC_1$.

## 5.2 Phase 2: Maximizing the Hammer Count

In the **second phase**, the attacker's goal is to maximize the target subbank's hammer count. As described in Section 4.2, the subbank hammers can be performed on any sequence of rows within the subbank, thereby modelling all possible RowHammer attacks to that subbank. To achieve this goal, the attacker should maximize the time it takes until Silver Bullet refreshes the victim row (where the attacker is attempting to induce RowHammer bit flips) in the target subbank. To this end, the attacker should aim to induce RowHammer bit flips in the subbank row (i.e., victim row) that was *last refreshed* by Silver Bullet. By doing so, the attacker can maximize the time it takes

Silver Bullet to refresh the victim row and therefore maximize the target subbank's hammer count. This is because Silver Bullet performs preventive refreshes on rows within a subbank in a round robin fashion, which means that Silver Bullet must first perform preventive refreshes on *every other row* in the target subbank before finally issuing a preventive refresh to the victim row (as shown in Figure 4). We refer to the time it takes Silver Bullet to perform preventive refreshes on every other row in the subbank as the *subbank refresh latency*.

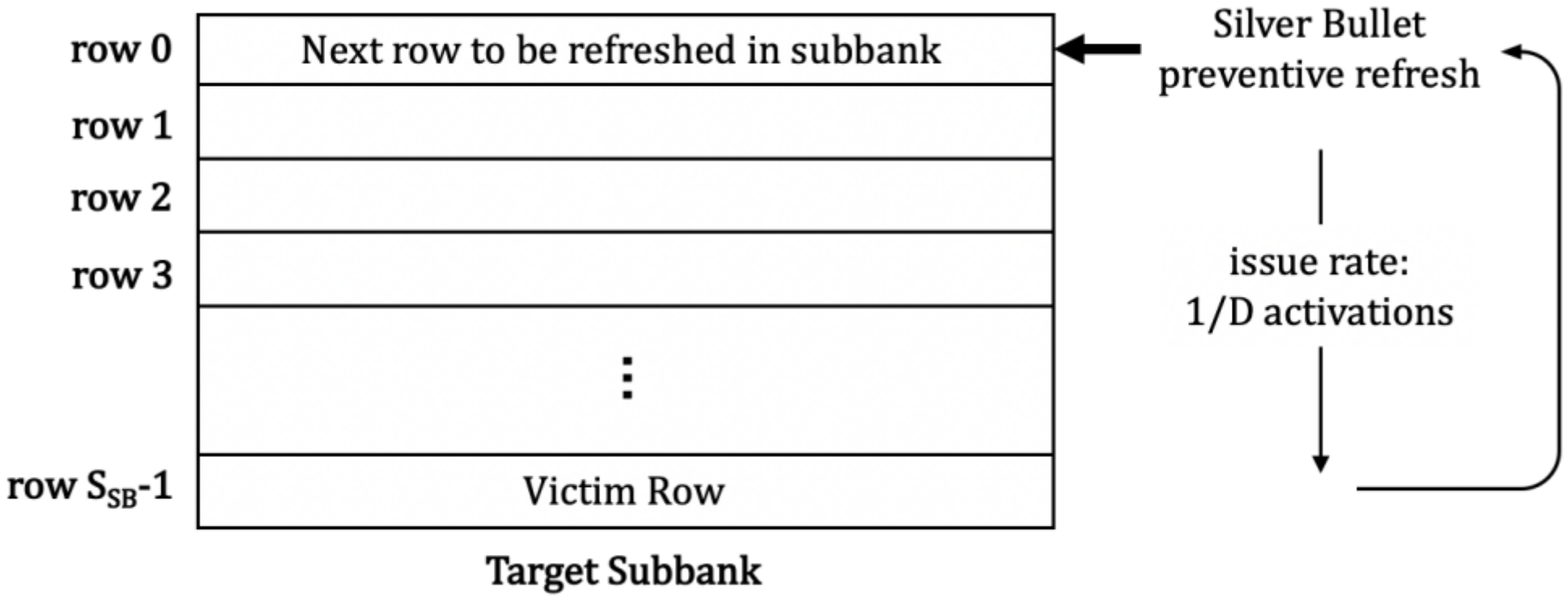

*Figure 4. Phase 2 of the attack. The time for Silver Bullet to perform preventive refresh operations on every row in the target subbank is found by multiplying the time to refresh a single row by the number of rows in the target subbank. During this time, an attacker can repeatedly hammer aggressor rows.*

Since Silver Bullet prioritizes preventive refreshes to subbanks with higher $PENDING$ values, Silver Bullet performs only a fraction of the preventive refreshes to the target subbank during this phase of the attack. This is because the attack is designed such that the $PENDING$ value of the target subbank remains below the highest $PENDING$ value across all subbanks, at least for a considerable amount of time during the attack). When Silver Bullet prioritizes preventive refreshes to another subbank, the subbank refresh latency of the target subbank increases and the attacker is given more time to hammer aggressor rows in the target subbank within a preventive refresh window.

We analyze Silver Bullet's security guarantees by assuming the worst case in Phase 2, where there exists an infinite number of subbanks with the highest $PENDING$ value ($P_{p1max}$) that a subbank can reach in Phase 1. With this worst-case assumption, we calculate the theoretical upper bound for the maximum number of hammers that an attacker can issue to the target subbank by the end of Phase 2. Therefore, any RowHammer-safe operating point that we derive with this assumption also holds for any subbank count and $PENDING$ value distribution across subbanks that Phase 1 results in.

We calculate the total hammer count that the target subbank (i.e., the subbank that contains the victim/aggressor rows) can receive in Phase 2 ($HC_2$) by individually examining Silver Bullet's latencies of *producing (Section 5.2.1)* and *consuming (Section 5.2.2)* the necessary preventive refreshes to refresh all rows in the target subbank and the amount of activations that an attacker can inject within these latencies. Since the consumption rate of preventive refreshes is greater than or equal to their production rate (Observation 1), Silver Bullet consumes (i.e., performs) preventive refreshes as they are produced with a latency that depends on the timings of performing preventive refreshes (i.e., $T$ and $R$). To calculate $HC_2$, we first find a theoretical upper bound for the hammer count that the attacker can reach while Silver Bullet produces preventive refreshes ($HC_{2A}$) in Section 5.2.1. Second, we calculate the additional hammer count ($HC_{2B}$) that the attacker can achieve due to the delay between producing and consuming all required preventive refreshes, in Section 5.2.2.

### 5.2.1 Activations until All Required Preventive Refreshes are Produced

We assume the target subbank's $PENDING$ value is initially $P_{p1max}$ when calculating $HC_{2A}$, the hammer count an attacker can achieve while producing the preventive refreshes required to refresh the entire target subbank. Due to our worst-case assumptions that 1) there are an infinite number of subbanks with a $PENDING$ value of $P_{p1max}$ and 2) Silver Bullet always prioritizes non-target subbanks that have the same $PENDING$ value as the target subbank (i.e., $P_{p1max}$)[6], Silver Bullet can only perform preventive refreshes to the target subbank when the target subbank's $PENDING$ value *exceeds* $P_{p1max}$. Therefore, in order for Silver Bullet to refresh the entire target subbank, $S_{SB}$ additional preventive refreshes must be produced or enqueued for the target subbank in Phase 2. Given that $S_{SB}$ preventive refreshes must be produced for the target subbank, and it takes $D$ hammers to produce a preventive refresh for each row, $HC_{2A}$ is simply a multiplication of $D$ and $S_{SB}$:

$$HC_{2A} = D.S_{SB}\ , D\ \geq 2(T/R + 1) \ldots\ldots\ldots\ldots\ldots\ldots\ldots\ldots\ldots\ldots\ldots (Expression\ 15)$$

**Observation 5.** The maximum hammer count in the target subbank that the attacker can achieve while producing all required preventive refreshes to a target subbank is proportional to both parameter $D$ and the subbank size as shown in *Expression 15*.

### 5.2.2 Hammer Count until All Required Preventive Refreshes are Performed

We examine the delay between enqueuing a preventive refresh (i.e., incrementing $PENDING$) and performing a preventive refresh (i.e., decrementing $PENDING$) and its effect on the worst-case hammer count that an attacker can reach in Phase 2. We can see from Observations 1 and 5, that Silver Bullet consumes preventive refreshes at a rate equal to or greater than the rate they are produced. However, Silver Bullet can only perform a burst of R preventive refreshes after a time window of $T$ activations. In the worst case, the last preventive refresh that Silver Bullet must perform is produced at a time such that Silver Bullet must wait for an additional time window of $T$ activations, which allows the attacker to hammer an aggressor row $T$ more times. Therefore, we calculate $HC_{2B}$ as shown in Expression 16.

$$HC_{2B} = T \ldots\ldots\ldots\ldots\ldots\ldots\ldots\ldots (Expression\ 16)$$

[6] This prioritization scheme delays Silver Bullet from performing preventive refreshes on the target subbank as long as possible, enabling an attacker to perform more hammers on the target subbank.

### 5.2.3 Calculating the Total Hammer Count in Phase 2

We calculate the total hammer count that an attacker can reach in Phase 2 as the sum of $HC_{2A}$ and $HC_{2B}$ (shown in Expression 17).

$$HC_2 = HC_{2A} + HC_{2B}, where$$

$$HC_{2A} = D.S_{SB} \; from\; Expression\; 15\; and\; HC_{2B} = T, from\; Expression\; 16$$

$$HC_2 = D.\, S_{SB} + T \ldots\ldots\ldots\ldots\ldots\ldots\ldots\ldots\ldots\ldots\ldots (Expression\; 17)$$

## 5.3 Calculating the Attack's Highest Hammer Count

We calculate the total hammer count ($HC_{attack}$) an attacker can reach during Phases 1 and 2 by summing the hammer counts achievable in the two phases ($HC_1 + HC_2$).

$$\boldsymbol{HC_1 = D.P_{p1max},} \qquad \boldsymbol{from\; Expression\; 14}$$

$$\boldsymbol{P_{p1max} = log_2(N_{SB})} \qquad \boldsymbol{from\; Expression\; 11}$$

$$\boldsymbol{HC_2 = D\, .S_{SB} + T} \qquad \boldsymbol{from\; Expression\; 17}$$

**Then,** $\boldsymbol{HC_{attack} = HC_1 + HC_2}$

$$\boldsymbol{HC_{attack} = (log_2(N_{SB}) + S_{SB})\, .D + T \ldots\ldots\ldots\ldots\ldots\ldots\ldots\ldots (Expression\; 18)}$$

***Note on Appendix A:*** *Appendix A evaluates the* ***same*** *upper bound as Expression 18 for the total hammer count that an attacker can reach by performing the Wave attack in Expression 18A. Thus, selecting a target subbank that was refreshed in Phase 1 does not affect our worst-case analysis.* This is due to our worst-case assumption regarding non-target subbanks' $PENDING$ values, where there are an infinite number of subbanks with a $PENDING$ value of $P_{p1max}$in Phase 2, which establishes a conservative upper bound for both cases. Therefore, we use Expression 18 for calculating the total hammer count an attacker can reach by performing the Wave attack.

## 5.4 Accounting for the Contribution of Periodic Refreshes to Total Hammer Count

We analyze the contribution of periodic DRAM-standard refresh operations to the total hammer count by calculating the number of periodic refresh operations that can be performed on aggressor rows (i.e., rows within the blast radius of the victim row) during an attack. Note that the attacker needs to complete the attack within a periodic refresh window. In the worst case the attack starts immediately after the victim row is refreshed. Thus, all aggressor rows are refreshed once until the time the victim row is refreshed again. Therefore, in the worst case periodic refreshes increase the total hammer count as much as the number of aggressor rows (2B). We calculate the hammer count due to periodic refreshes ($HC_{ref}$) using Expression 19.

$$HC_{ref} = 2B \ldots\ldots\ldots\ldots\ldots\ldots\ldots\ldots\ldots\ldots\ldots\ldots\ldots (Expression\; 19)$$

## 5.5 Total Hammer Count

We calculate the total hammer count as the sum of 1) the hammer count that the attacker can reach by performing row activations and 2) the effective hammer count caused by periodic refreshes during the attack, as shown in Expression 20.

$$HC_{total} = HC_{attack} + HC_{ref}$$

$$\boldsymbol{HC_{attack}} = (\boldsymbol{log_2(N_{SB}) + S_{SB}}) \,.\, \boldsymbol{D + T} \quad \boldsymbol{from\ Expression\ 18}$$

$$\boldsymbol{HC_{ref}} = 2B \qquad from\ Expression\ 19$$

$$\boldsymbol{HC_{total}} = (\boldsymbol{log_2(N_{SB}) + S_{SB}}) \,.\, \boldsymbol{D + T + 2B} \ldots\ldots\ldots\ldots\ldots\ldots\ldots (Expression\ 20)$$

## 5.6 Key Takeaways on the Worst-Case RowHammer Attack

**In this section, we have provided a formalized proof of the worst-case access pattern that an attacker can issue to a Silver Bullet-protected system to maximize the Hammer Count that a subbank can receive within a preventive refresh window. We next analyze the design constraints that Silver Bullet must satisfy in order to guarantee RowHammer-safe operation.**

# 6. Conditions for RowHammer-Safe Operation

From our calculations in Section 5, an attacker can reach up to a $HC_{total}$ hammer count during a RowHammer attack on a Silver Bullet-protected system, where $HC_{total}$is calculated using Expression 20. Silver Bullet can securely prevent RowHammer bit flips on any DRAM chip with a $UHC_{DRAM}$ value greater than $HC_{total}$, i.e.,

$$UHC_{DRAM} > HC_{total}, where HC_{total} = (\boldsymbol{log_2(N_{SB}) + S_{SB}}) \,.\, \boldsymbol{D + T + 2B} \quad from\ Expression\ 20$$

$$UHC_{DRAM} > \boldsymbol{D}.\,(\boldsymbol{log_2(N_{SB}) + S_{SB}}) + \boldsymbol{T + 2B} \ldots\ldots\ldots\ldots\ldots (\boldsymbol{Expression\ 21})$$

For a Silver Bullet-protected system to ensure RowHammer-safe operation, Silver Bullet should be configured in such a way that the unsafe hammer count of a DRAM chip ($UHC_{DRAM}$) should be larger than the maximum RowHammer-safe hammer count that Silver Bullet supports, which we call the *Tolerable Hammer Count* ($\boldsymbol{THC}$). $THC$is thus defined to be the maximum hammer count that a worst-case attack can ever reach, as follows:

$$THC = D.\,(\boldsymbol{log_2(N_{SB}) + S_{SB}}) + T + \boldsymbol{2B} \ldots\ldots\ldots\ldots\ldots\ldots\ldots (Expression\ 22)$$

## 6.1 Putting all Constraints Together

To prevent RowHammer bit flips in a system with Silver Bullet, the designer should configure Silver Bullet's parameters $D$, $S_{SB}$,and $N_{SB}$such that *Expression 21* is satisfied for a given DRAM chip's reliability characteristics ($UHC_{DRAM}$ and $B$) and limitations imposed by the DRAM interface protocol ($R$ and $T$). We revisit each of $D$, $S_{SB}$, and $N_{SB}$ here, briefly:

- Parameter $D$ is an integer value within the range of $[2(T/R+1), UHC_{DRAM})$ from *Expression 1*.
- Subbank size ($S_{SB}$) is an integer value within the range $[2B, S_B]$, where $S_B$ is the number of rows in a bank. If subbanks are not equally sized, the attacker should target the largest subbank in order to maximize the hammer count to the target subbank.[7] In this case, $S_{SB}$ is the size of the largest subbank.
- $N_{SB}$ is the number of subbanks, which depends on bank size ($S_B$) and the configuration of subbank sizes. In our analysis we assume that all subbanks are equally sized. Therefore, $N_{SB} = S_B/S_{SB}$.

**Observation 6.** Silver Bullet guarantees RowHammer-safe operation for a DRAM chip whose $UHC_{DRAM}$ value is larger than the maximum tolerable hammer count value ($THC$) Silver Bullet supports. To summarize, as explained above:

$$UHC_{DRAM} > THC = D.(log_2(N_{SB}) + S_{SB}) + T + 2B, where$$

- $D \geq 2(T/R+1)$,
- $S_B \geq S_{SB} \geq 2B$
- $N_{SB} = S_B/S_{SB}$

## 6.2 Key Takeaways on Silver Bullet Design Constraints

**We conclude that Silver Bullet prevents all RowHammer attacks when configured correctly given system characteristics: $UHC_{DRAM}, B, R, and\ T$ as stated in Observation 6.**

**Based on our proof, we conclude that Silver Bullet is a promising RowHammer prevention mechanism that can be configured to operate securely against RowHammer attacks and scales to prevent RowHammer bit flips in DRAM chips with relatively small unsafe hammer count values (e.g., 1000).**

# 7. Silver Bullet Table Size

As described in Section 2.1, the Silver Bullet table consists of one entry per subbank where each entry is comprised of three fields: $FRAC$, $PENDING$, and $LOCAL_INDEX$. Since the $FRAC$ field must support values between 0 and D, it should be $\lceil log(D) \rceil$ bits wide. The maximum value of the $PENDING$ field during Phase 1 is defined as $\boldsymbol{P_{p1max}}$. As we explain in Section 5.2.1, Silver Bullet prioritizes preventive refreshes on a subbank when the subbank's $PENDING$ value exceeds $P_{p1max}$. In the worst case, Silver Bullet is unable to perform any preventive refresh for T activations, which allows a subbank's $PENDING$ value to increase to up to $P_{p1max} + (T+R)/D$. From Expression 1, $D \geq 2(T/R+1)$ and thus $(T+R)/D \leq R/2$. Therefore, the $PENDING$ value can reach up to $\boldsymbol{P_{p1max}} + \boldsymbol{R/2}$ in Phase 2. To support these values, the $PENDING$ field should be $\lceil log_2(P_{p1max} + R/2)\rceil$ bits wide. From Expression 11, $\boldsymbol{P_{p1max}}$ *cannot* be larger than $log_2(N_{SB})$. After substituting $log_2(N_{SB})$ for $P_{p1max}$, we determine that the $PENDING$ field should be $\lceil log_2(log_2(N_{SB}) + R/2)\rceil$ bits wide. Since the $LOCAL_INDEX$ field must be able to reference any row in a subbank, it should be $\lceil log(S_{SB})\rceil$bits wide. To find the total size of the Silver Bullet table per bank, we multiply the sum of these values by the number of subbanks in a bank $(N_{SB})$, as follows.

[7] In a realistic design, the subbank size may correspond to the subbank's degree of RowHammer vulnerability. Therefore, targeting the largest subbank to attack or selecting the subbank to target solely based on size may not necessarily be the best method for the worst-case attack. We leave the construction of a worst-case attack in the presence of varying subbank sizes and vulnerabilities to future work.

$$table_{size} = \left(\lceil log_2(D)\rceil + \left\lceil log_2\left(log_2(N_{SB}) + \frac{R}{2}\right)\right\rceil + \lceil log_2(S_{SB})\rceil\right).N_{SB} \ (Expression\ 23)$$

We assume that Silver Bullet maintains an SRAM table in the DRAM chip. Based on prior works' SRAM table overhead analysis on a DRAM chip [6, 7], we estimate that an SRAM cell requires 20x the area overhead of a DRAM cell. In our area overhead estimations in Section 9, we conservatively assume that the area overhead of an SRAM cell equals the area overhead of 200 DRAM cells.

To minimize the Silver Bullet table size, one table entry can be used to track hammers across multiple subbanks if Silver Bullet can simultaneously refresh different rows in all subbanks that share the same table entry. In this implementation, the FRAC counter in the table entry is incremented each time a hammer is performed on any of the subbanks that share the entry. When Silver Bullet performs a preventive refresh, the refresh operation is simultaneously executed in all subbanks sharing the entry. If Silver Bullet can be configured such that a single table entry can be shared across $n$ subbanks, the overall table size can be reduced by a factor of $n$.

# 8. Summary of Key Parameters Used in Security Analysis

Table 3 lists key parameters that we use in modeling the worst case attack and calculating the maximum tolerable hammer count that a Silver Bullet implementation supports. We also include a table size calculation.

*Table 3. Key parameters demonstrating Silver Bullet's behavior.*

| | Expression | Definition |
|---|---|---|
| $\boldsymbol{k}$ | $0 < k = \frac{2T - R}{2T + (D-1)\,R} \leq \frac{D-1}{2D-1}$ | The reduction factor of subbank counts across consecutive iterations of the first phase of the worst-case attack against Silver Bullet |
| $\boldsymbol{P_{p1max}}$ | $P_{p1max} = i_{last} = log_k\ (1/N_{SB})$ <br> Theoretical upper bound for $P_{p1max} = i_{last}\ is\ log_2(N_{SB})$ | Maximum $PENDING$ value the attacker can reach across all subbanks in Phase 1. $P_{p1max}$ equals $i_{last}$ (i.e., the number of iterations in Phase 1) because $PENDING$ is incremented by one in each iteration |
| $\boldsymbol{HC_1}$ | $D.P_{p1max}$ | The maximum number of times an aggressor row can be hammered in Phase 1 of the attack |
| $\boldsymbol{HC_2}$ | $D.(P_{p1max} + S_{SB})$ | The maximum number of times an aggressor row can be hammered in Phase 2 of the attack |
| $\boldsymbol{HC_{total}}$ | $D.(P_{p1max} + S_{SB})$ <br> $+T + 2B$ | The total hammer count that an attacker can achieve by performing the worst-case attack. This is the sum of $HC_1$ and $HC_2$ |
| $\boldsymbol{THC}$ | $D.(log_2(N_{SB}) + S_{SB})$ <br> $+T + 2B$ | **Maximum hammer count that can be tolerated by a Silver Bullet implementation, given the worst-case attack** |
| $TableEn$ | $\lceil log_2(D)\rceil + \lceil log_2(log_2(N_{SB}) + R/2)\rceil + \lceil log_2(S_{SB})\rceil$ | The size of each Silver Bullet table entry in bits |
| $Table\ Size$ | $(Table\ Entry\ Size)\,.\,N_{SB}$ | The size of the Silver Bullet table in bits |

# 9. Finding RowHammer-Safe Silver Bullet Design Points

To demonstrate Silver Bullet's flexibility in implementation, we identify design points that are free of RowHammer-induced bitflips and corresponding Silver Bullet configuration parameters for a given set of $THC$ values. These parameters can be chosen to prevent RowHammer bit flips in any DRAM chip where $UHC_{DRAM} > THC$. Figure 5 plots the RowHammer-safe operating points that we calculate from the constraints and inequalities presented in Observation 6 for a given $THC$. The x-axis shows the minimum supported $THC$, while the y-axis shows the additional refreshes required per regular row activation (calculated by $1/D$). Different color and marker combinations show different subbank size configurations. We report the number of Silver Bullet table entries (described in Section 2.1) per 1024 rows for each subbank size.

## 9.1 Supported THC Values by Various Silver Bullet Configurations

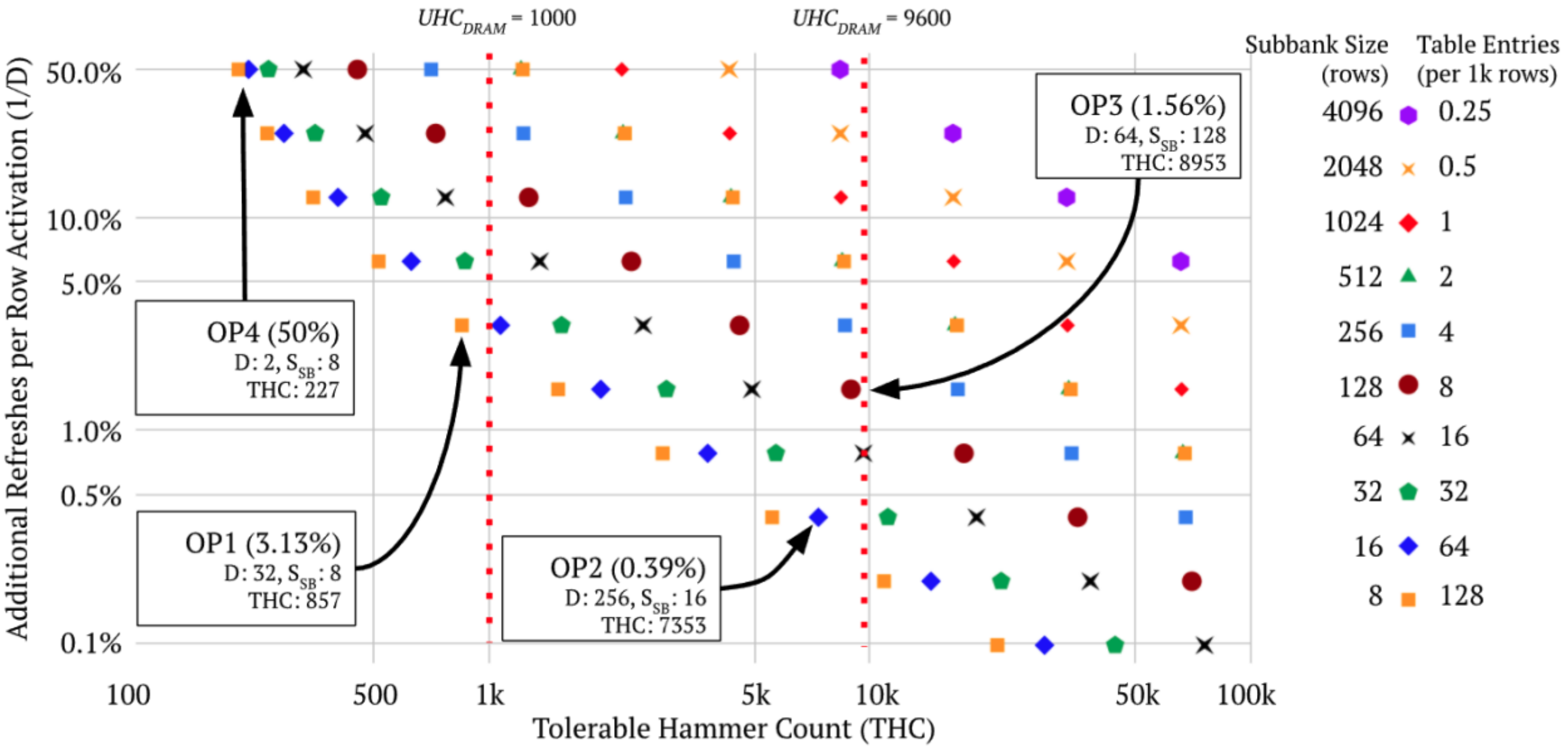

***Figure 5. RowHammer-safe operating points for varying THC values (B=4). Operating points supporting a smaller THC value are more desirable, all else being equal, since they provide better protection against RowHammer attacks that can flip bits with fewer hammer counts. OP1-4 are highlighted as example operating points and are discussed in the text.***

In this analysis, we assume a realistic DRAM bank size ($N_B$) of 64k rows, a blast radius ($B$) of 4 rows, and a realistic $T$ value of 177 activations.[8] We identify the operating points by evaluating a wide range of Silver Bullet configurations. We vary the subbank size ($S_{SB}$) between 8 and 4096 rows, assume that all subbanks are equally sized, and sweep parameter $D$ between 2 and 1024. We make three key observations from Figure 5.

[8] Assuming (1) a time interval of 7.8us between two REF commands ($t_{REFI}$=7.8us), (2) a minimum time delay of 46ns between two consecutive activations targeting the same bank ($t_{RC}$=46ns), and (3) a refresh latency of 350ns ($t_{RFC}$=350ns), we calculate T as 177 activations, i.e., ($t_{REFI}$+$t_{RFC}$)/$t_{RC}$ .

- **Silver Bullet supports very low $THC$ values (i.e., <1000), depending on the configuration of its parameters.** For example, at 3.13 additional refreshes per 100 hammers and using a subbank size of 8 rows, the minimum $THC$ value is 857, as shown by OP1 in Figure 5. The Silver Bullet table size for OP1 is 13KB (in SRAM cells) based on Expression 23.

  Today, the lowest activation count for a successful RowHammer attack reported in the literature is 4800 row activations for each aggressor row in a double-sided attack (i.e., alternate activations to two rows that are physically adjacent to the victim row) [**3**]. This equates to 9600 subbank activations ($UHC_{DRAM}$) on a Silver Bullet protected system because there is always at least one subbank that contains both aggressor rows of a double-sided attack whose activation counter ($FRAC$) increments when either of the aggressor rows activated. Therefore, a Silver Bullet implementation with a $THC$ value lower than 9600 (i.e., any operating point to the left of the dashed red line) prevents RowHammer bit flips on any tested chip so far. We highlight two sample RowHammer-safe operating points (OP2 and OP3) in Figure 5 to demonstrate the design tradeoffs. One Silver Bullet implementation, operating at OP2, requires a subbank size of 16 rows (i.e., 64 Silver Bullet table entries per 1k rows), which translates to a Silver Bullet table size of 8KB, and performs 0.39 preventive refreshes per 100 row activations. Another Silver Bullet implementation, operating at OP3, requires a subbank size of 128 rows (i.e., only 8 Silver Bullet table entries per 1k rows), which translates to a significantly smaller Silver Bullet table size of 1.06KB (or area equivalent to 207.51KB of DRAM capacity), but performs 1.56 preventive refreshes per 100 row activations.

  Using Figure 5 (or a plot similarly derived from our mathematical analysis and resulting equations), one can choose RowHammer-safe Silver Bullet operating points and parameters with different efficiency-area tradeoffs.

- As the RowHammer vulnerability of a DRAM device increases ($UHC_{DRAM}$ decreases), Silver Bullet must be implemented with a lower $THC$ design point (where $THC < UHC_{DRAM}$). Supporting lower $THC$ values requires a larger number of additional preventive refreshes, as Figure 5 demonstrates.

- Based on Observation 6, Silver Bullet can reduce the number of preventive refreshes performed to the target subbank by reducing the subbank size. However, this subbank size reduction comes at the cost of increasing the number of subbanks in a DRAM bank, and thus the area overhead of Silver Bullet (which is proportional to the number of entries in the Silver Bullet table). Therefore, efficiency and area overhead can be traded-off, as Figure 5 demonstrates.

- Given the parameters we evaluate, the most extreme operating point that provides the lowest $THC$, shown as OP4, leads to a $THC$ value of 227. Clearly, this operating point is possible at high overheads (a very large 50 additional refreshes per 100 activations and a very small subbank size of 8 rows, i.e., 11KB of Silver Bullet table size..

## 9.2 Supported THC Values by Various Silver Bullet Configurations when B=1

Figure 6 repeats the same analysis of RowHammer-safe operating points when blast radius is 1, as opposed to 4. In this analysis, we use a fixed DRAM bank size $(N_B)$ of $64k$ and a $T$ value of 177 (as we did for Figure 5 in Section 9.1). We vary the subbank size $(S_{SB})$ between 2 and 4096 rows (assuming equally-sized subbanks), and sweep parameter $D$ between 2 and 1024.

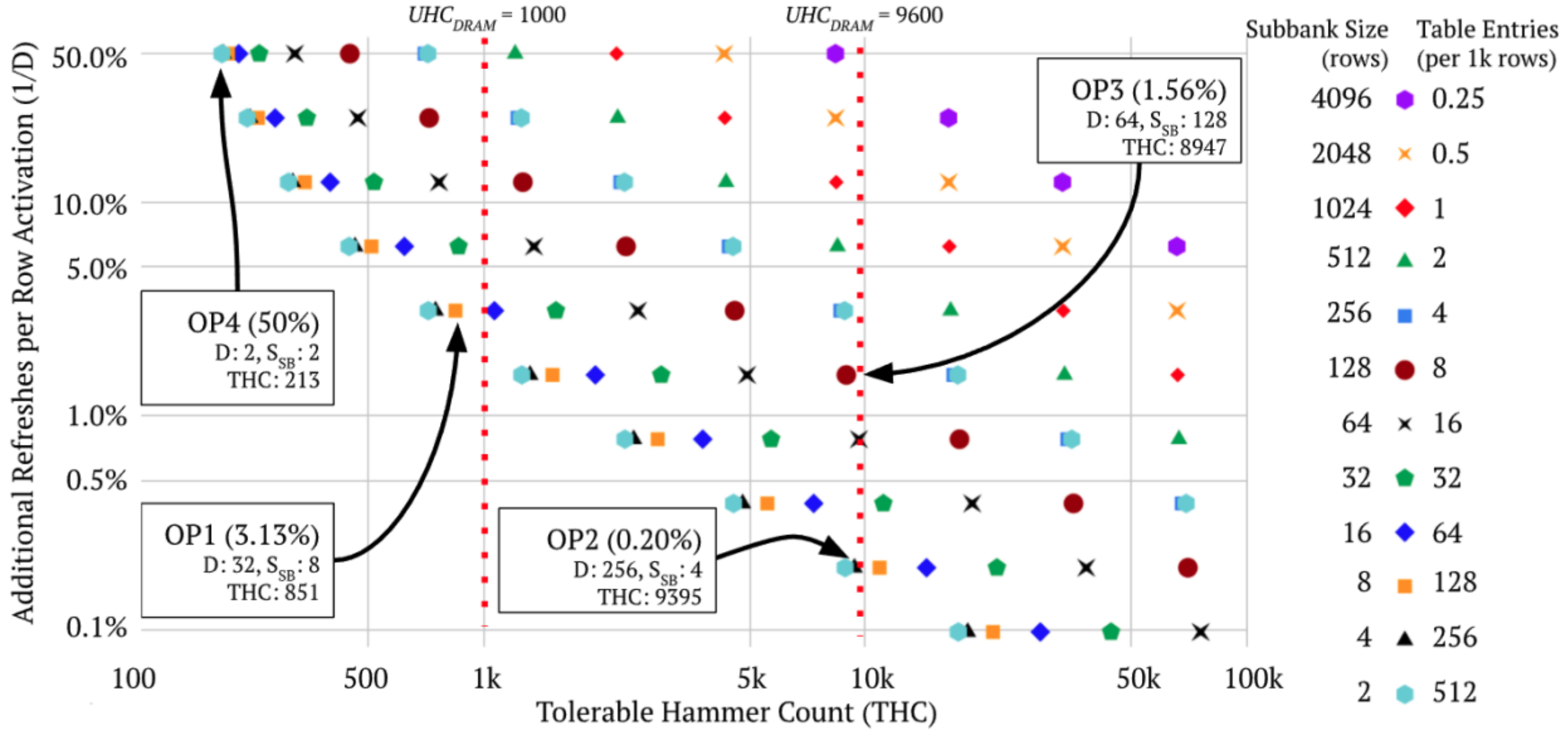

*Figure 6. RowHammer-safe operating points for varying THC values (B=1).*

We make three key observations from Figure 6, comparing it to Figure 5:

First, the subbank size and additional refresh rate of highlighted RowHammer-safe operating points OP1 and OP3, do not change from Figure 5 (B=4) to Figure 6 (B=1). However, tolerable hammer count values of these points reduce from 857 to 851 for OP1 and from 8953 to 8947 for OP3, showing that a lower blast radius strengthens the RowHammer-free operation capability of Silver Bullet.

Second, we observe that a smaller blast radius enables Silver Bullet to use smaller subbank sizes, which enable new RowHammer-safe operating points with a smaller number of required additional refreshes per row activation. OP2 in Figure 6 uses a subbank size of 4 rows, which is not possible when the blast radius is 4 (in Figure 5), since the subbank size must be greater than or equal to double the blast radius (from Observation 6). Compared to OP2 in Figure 5, OP2 in Figure 6 reduces the required additional refreshes per row activation from 0.39% to 0.20% by using a significantly smaller subbank size of 4 rows versus 16 rows, which increases the Silver Bullet table size from 8KB to 28KB..

Third, we observe that the minimum THC observed with the used parameters goes down to as low as 213, at OP4. Clearly, this operating point is possible at high overheads (a very large 50 additional refreshes per 100 activations and a very small subbank size of 2 rows, which translates to a Silver Bullet table size of 36KB.

# 10. Sensitivity Analyses

In this section, we evaluate Silver Bullet with various configuration parameter values and investigate the impact of different parameters on the minimum tolerable hammer count ($THC$) that Silver Bullet can guarantee RowHammer-free operation for.

## 10.1 Sensitivity of Tolerable Hammer Count to Parameter D and Subbank Size

Observation 6 shows that the lower bound of $THC$ is strongly affected by both $D$ and subbank size. In Figure 7, we plot the effects of D and subbank size on THC, where D is on the x-axis and different curves represent different subbank sizes. This analysis assumes a realistic DRAM chip with 64k rows per bank ($S_B$), a RowHammer blast radius ($B$) of 4, and a T value of 177.

We make three observations from Figure 7.

First, we observe that reducing parameter $D$ enables Silver Bullet to ensure RowHammer-free operation at lower THC values. This is because $D$ is inversely proportional to the rate at which Silver Bullet performs preventive refreshes on a target subbank, and a higher preventive refresh rate reduces the maximum hammer count that an attacker can induce on an aggressor row.

Second, we observe that the DRAM interface protocol can limit Silver Bullet (i.e., by limiting the minimum parameter $D$ value, as described in Observation 1, shown with dashed vertical lines).

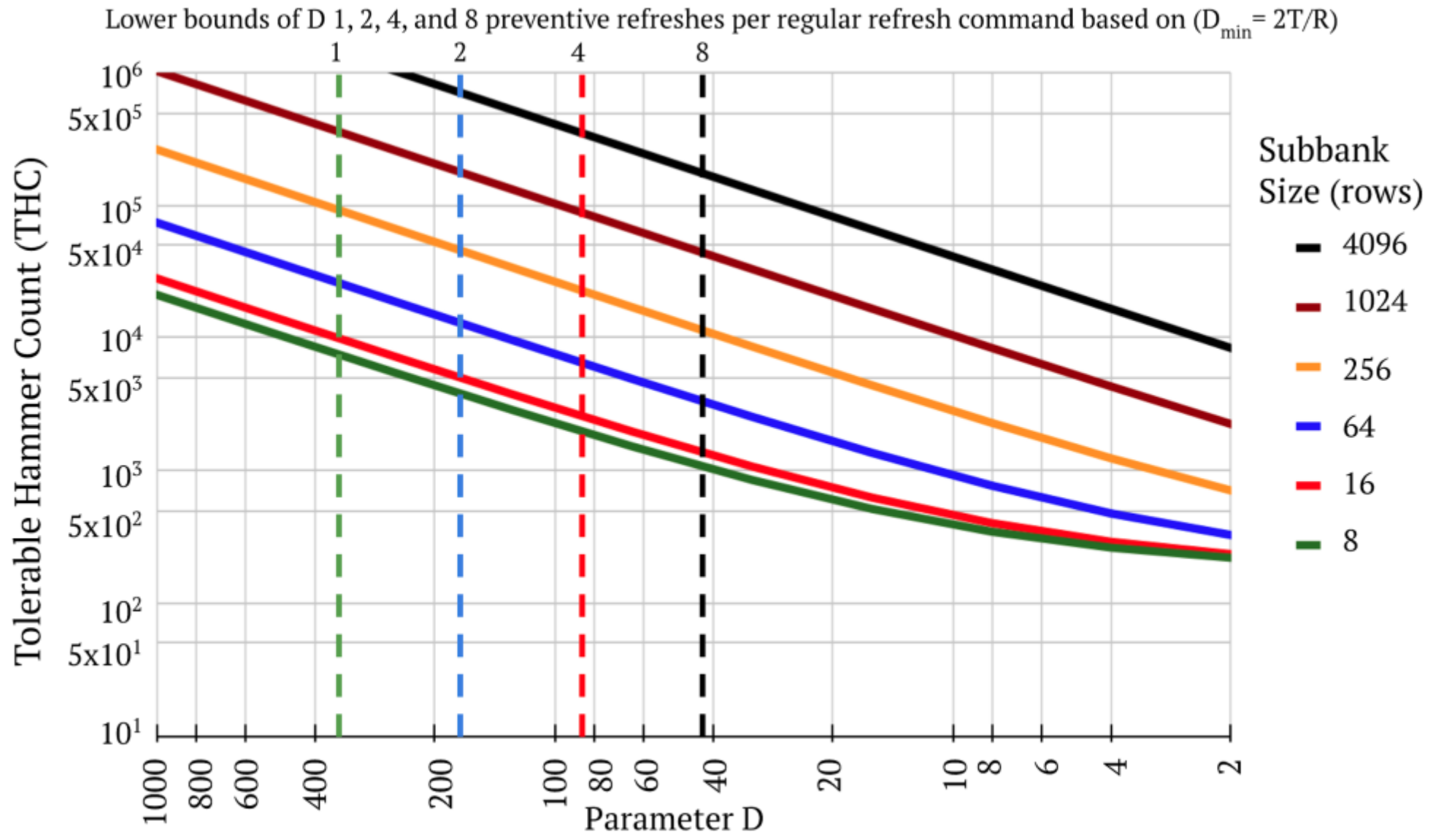

*Figure 7. Effect of Parameter D and subbank size on Silver Bullet's Tolerable Hammer Count (THC).*

Third, we plot the minimum parameter $D$ values using the following Silver Bullet/DRAM configurations: 1) Silver Bullet can perform preventive refreshes for (a) one, (b) two, (c) four, and (d) eight rows within the latency of a regular refresh command ($t_{RFC}$), which can be issued every 7.8$\mu s$($t_{REFI}$), and 2) we assume a minimum delay between two consecutive row activations targeting the same bank to be $46\,ns$ ($t_{RC}$). We observe that while Silver Bullet is capable of supporting very low tolerable hammer counts (e.g., <500), the DRAM interface protocol can limit the devices on which Silver Bullet can prevent RowHammer bit flips due to such low hammer counts. Note that this limitation can be avoided with simple modifications to the DRAM interface protocol (e.g., DRAM Initiated Pause) or by implementing Silver Bullet in the memory controller.

## 10.2 Sensitivity of Tolerable Hammer Count to DRAM Bank and Subbank Sizes

Figure 8 plots the minimum supported $THC$ for a range of DRAM bank and subbank sizes. In this study, we assume a blast radius of 4 and a T value of 177. We independently sweep bank size (Figure 8a) and subbank size (Figure 8b) to observe the isolated effects of each parameter for four different parameter D values that correspond to 0.78%, 1.56%, 3.13%, and 6.25% additional refreshes per activation. We sweep bank size from 16k rows to 512k rows with a constant subbank size of 64 rows. Similarly, we sweep subbank size from 8 rows to 8k rows with a constant bank size of 64k rows. We calculate the maximum supported $THC$ value by using *Expression 22*.

We observe that the subbank size greatly affects the THC (Figure 8b), whereas the DRAM bank size does *not* (Figure 8a). This is because reducing the subbank size ($S_{SB}$) reduces the hammer count the attacker can reach in Phase 2 ($HC_2$), but it also increases the number of subbanks ($N_{SB}$) and thus the hammer count the attacker can reach in Phase 1 ($HC_1$), leading to a tradeoff between $S_{SB}$ and $N_{SB}$ when maximizing $THC$ (i.e., $HC_1 + HC_2$). Since $N_{SB}$ contributes logarithmically to $THC$, and $S_{SB}$contributes linearly to $THC$, Silver Bullet's $THC$ is significantly higher (i.e., worse) when it is implemented with larger subbank sizes. Similarly, increasing the bank size with a fixed $S_{SB}$ only increases $N_{SB}$, which contributes logarithmically to $THC$ (as observed in Figure 8a).

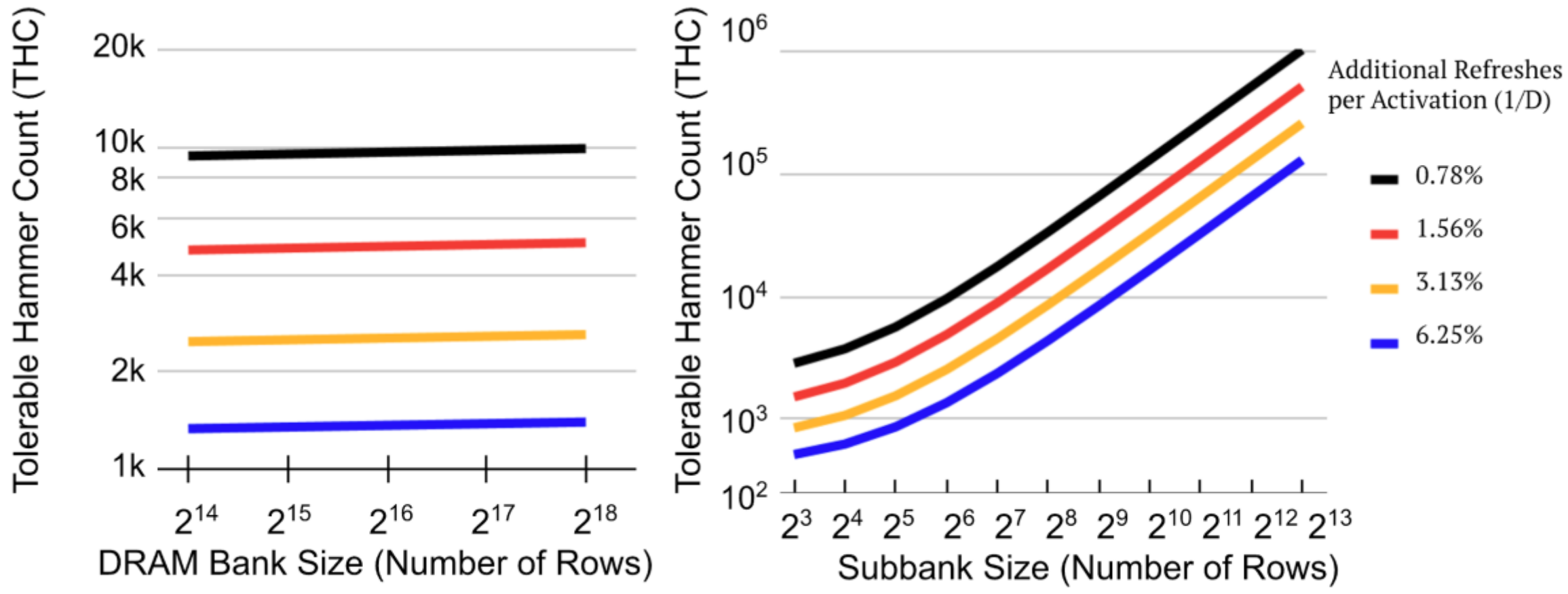

a) THC vs Bank Size (Subbank size: 64 rows) b) THC vs Subbank Size (Bank size: 64k rows)

***Figure 8. Sensitivity of THC to DRAM Bank Size and Subbank Size.*** **We assume *B*=4.**

## 10.3 Sensitivity of Tolerable Hammer Count to Blast Radius

Figure 9 plots the sensitivity of the minimum supported $THC$ value to different blast radius values. We sweep blast radius from 1 to 8 (the maximum observed Blast Radius [3]) for eight different subbank sizes plotted in different colors (indicated by the legend).

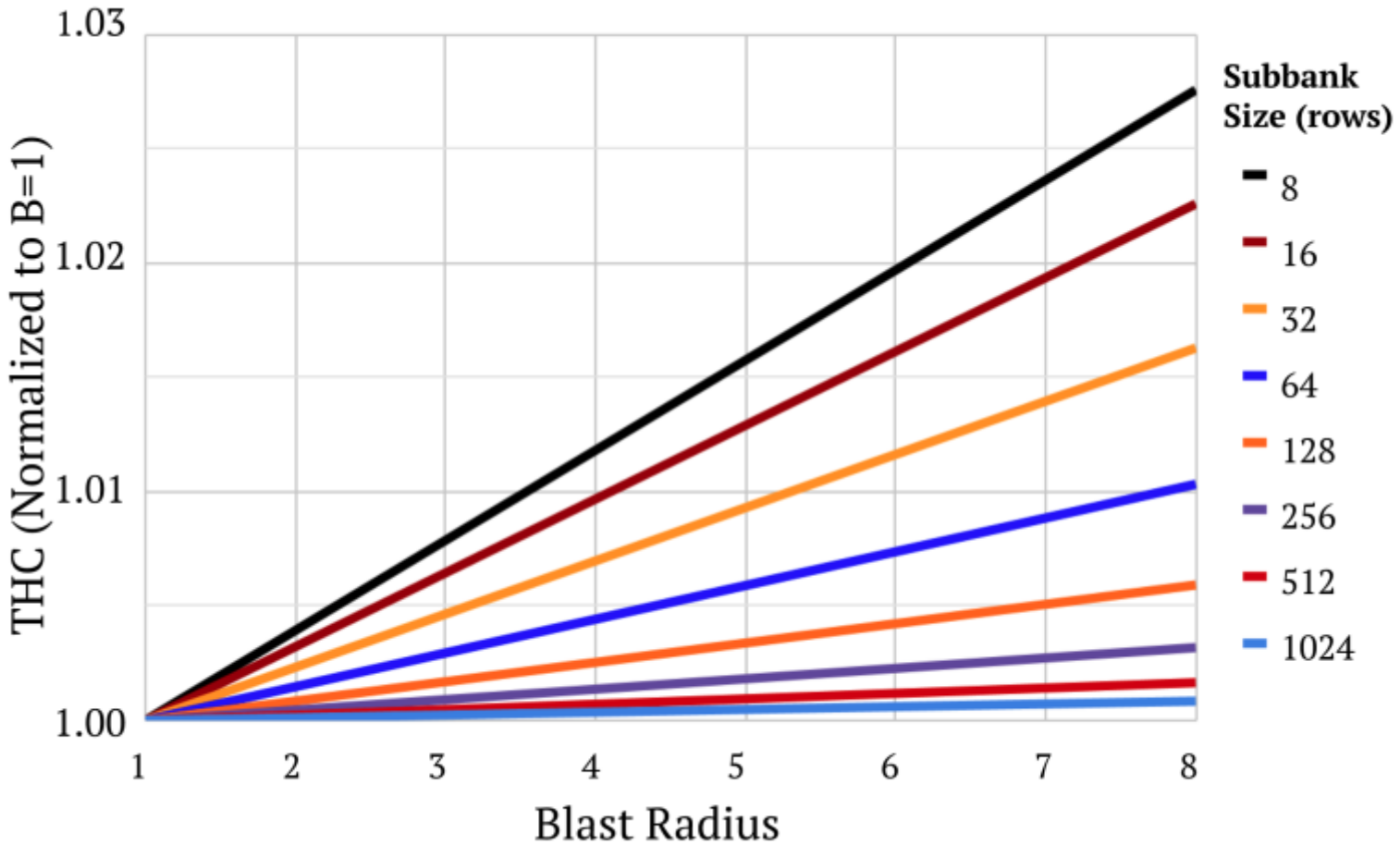

***Figure 9. Effect of blast radius on the minimum supported THC.***

We observe that a larger blast radius (B) value leads to a marginal increase in the $THC$ value (i.e., less than 3%). This effect reduces as the subbank size increases. Therefore, we conclude that Blast Radius does not significantly affect Silver Bullet's security guarantees for a given subbank size, while Blast Radius defines a lower bound for the subbank size based on Observation 6.

# 11. Conclusion

We provide a mathematical proof that shows that the Silver Bullet algorithm, when configured and implemented properly, securely prevents RowHammer attacks. Our study concludes that Silver Bullet is a promising RowHammer prevention mechanism that can be configured to operate securely against RowHammer attacks at various efficiency-area tradeoff points, supporting relatively small tolerable hammer count values (e.g., 1000) and Silver Bullet table sizes (e.g., 1.06KB).

## References

[1] JEDEC Solid State Technology Association, “DDR4 SDRAM Specification,” 2017.

[2] E. Lee, I. Kang, S. Lee, G. Edward Suh, and J. Ho Ahn, “TWiCe: Preventing Row-Hammering by Exploiting Time Window Counters,” in ISCA, 2019.

[3] J. S. Kim, M. Patel, A. G. Yaglikci, H. Hassan, R. Azizi, L. Orosa, and O. Mutlu, “Revisiting RowHammer: An Experimental Analysis of Modern Devices and Mitigation Techniques," in ISCA 2020.

[4] JEDEC Standard High Bandwidth Memory (HBM) DRAM Specification JESD235A, 2015.

[5] JEDEC Standard DDR5 SDRAM Specification, JESD79-5 2020.

[6] D. Lee, Y. Kim, V. Seshadri, J. Liu, L. Subramanian, and O. Mutlu, “Tiered-Latency DRAM: A Low Latency and Low Cost DRAM Architecture,” in HPCA 2013.

[7] S. Thoziyoor, J. H. Ahn, M. Monchiero, J. Brockman, and N. Jouppi, “A Comprehensive Memory Modeling Tool and Its Application to the Design and Analysis of Future Memory Hierarchies,” in ISCA 2008.

[8] Y. Kim, R. Daly, J. Kim, C. Fallin, J. H. Lee, D. Lee, C. Wilkerson, K. Lai, and O. Mutlu. “Flipping Bits in Memory without Accessing them: An Experimental Study of DRAM Disturbance Errors,” ISCA 2014.

[9] P. Frigo, E. Vannacci, H. Hassan, V. van der Veen, O. Mutlu, Cristiano Giuffrida, Herbert Bos, and Kaveh Razavi. “TRRespass: Exploiting the Many Sides of Target Row Refresh,” S&P 2020.

[10] F. Devaux and R. Ayrignac. “Method and Circuit for Protecting a DRAM Memory Device from the Row Hammer Effect,” US Patent 10,885,966. 2021. http://patft.uspto.gov/netacgi/nph-Parser?Sect1=PTO2&Sect2=HITOFF&p=1&u=%2Fnetahtml%2FPTO%2Fsearch-bool.html&r=1&f=G&l=50&co1=AND&d=PTXT&s1=10885966&OS=10885966&RS=10885966

[11] O. Mutlu and J. Kim, “RowHammer: A Retrospective” TCAD Special Issue on Top Picks in Hardware and Embedded Security, 2019.

[12] Google Project Zero, “Exploiting the DRAM RowHammer Bug to Gain Kernel Privileges,” https://googleprojectzero.blogspot.com/2015/03/exploiting-dram-rowhammer-bug-to-gain.html, 2015.

[13] O. Mutlu, “The RowHammer Problem and Other Issues We May Face as Memory Becomes Denser” Invited Paper in DATE, 2017.

[14] A. G. Yaglikci, M. Patel, J. S. Kim, R. Azizi, A. Olgun, L. Orosa, H. Hassan, J. Park, K. Kanellopoulos, T. Shahroodi, S. Ghose, and O. Mutlu, “BlockHammer: Preventing RowHammer at Low Cost by Blacklisting Rapidly-Accessed DRAM Rows,” in HPCA, 2021.

[15] Y. Park, W. Kwon, E. Lee, T. J. Ham, J. Ho Ahn and J. W. Lee, “Graphene: Strong yet Lightweight Row Hammer Protection,” in MICRO 2020.

[16] K. K. Chang, P. J. Nair, S. Ghose, D. Lee, M. K. Qureshi, and O. Mutlu, “Low-Cost Inter-Linked Subarrays (LISA): Enabling Fast Inter-Subarray Data Movement in DRAM,” in HPCA 2016.

[17] Y. Kim, V. Seshadri, D. Lee, J. Liu, and O. Mutlu, “A Case for Exploiting Subarray-Level Parallelism (SALP) in DRAM,” in ISCA 2012.

[18] K. Chang, D. Lee, Z. Chishti, A. Alameldeen, C. Wilkerson, Y. Kim, and O. Mutlu, “Improving DRAM Performance by Parallelizing Refreshes with Accesses” in HPCA, 2014.

[19] T. Yang and X. -W. Lin,“Trap-Assisted DRAM RowHammer Effect,” EDL, 2019.

[20] K. Park, C. Lim, D. Yun, and S. Baeg, “Experiments and Root Cause Analysis for Active-Precharge Hammering Fault in DDR3 SDRAM under 3xnm Technology,” Microelectronics Reliability, 2016.

# APPENDIX A. Alternative Target Subbank Selections in Worst-Case Attack

This section calculates the total hammer count that an attacker can achieve if the attacker targets a subbank that was preventively refreshed in Phase 1. We provide alternatives to Sections 5.2.1, 5.2.3, and 5.3 under this assumption.

## 5.2.1A Hammer Count until all Required Preventive Refreshes are Produced

We analyze the initial state of the target subbank in Phase 2, where Silver Bullet performs a preventive refresh on the target subbank when the target subbank's $PENDING$ value is $P_{REF}$ during Phase 1, where $1 \leq P_{REF} \leq P_{p1max}$ (Expression 12). From this initial state, we calculate $HC_{2A}$, the hammer count an attacker can achieve while producing the preventive refreshes required to refresh the entire target subbank.

Due to our worst-case assumptions that 1) there are an infinite number of subbanks with a $PENDING$ value of $P_{p1max}$ and 2) Silver Bullet always prioritizes non-target subbanks that have the same $PENDING$ value as the target subbank (i.e., $P_{p1max}$)[9], Silver Bullet can only perform preventive refreshes to the target subbank when the target subbank's $PENDING$ value *exceeds* $P_{p1max}$. If an attacker targets a subbank that was refreshed by Silver Bullet when its $PENDING$ value was $P_{REF}$ in Phase 1, the target subbank's $PENDING$ value would be decremented to $P_{REF} - 1$, and the number of preventive refreshes required for the entire target subbank to be refreshed in Phase 2 becomes $S_{SB} - 1$, based on our analysis for *Expression 13*. In Phase 2-A, the attacker first hammers the target subbank until the target subbank's $PENDING$ value reaches $P_{p1max}$ from $P_{REF} - 1$, which takes $(P_{p1max} - (P_{REF} - 1))\,.\,D$ hammers. From this point, since Silver Bullet has already performed one preventive refresh on the target subbank in Phase 1, $S_{SB} - 1$ additional preventive refreshes must be produced or enqueued for the target subbank. Given that $S_{SB} - 1$ preventive refreshes must be produced for the target subbank, and it takes $D$ hammers to produce a preventive refresh for each row, the achievable hammer count is simply a multiplication of $D$ and $S_{SB} - 1$.

To calculate $HC_{2A}$, we sum these individual terms:

$$
\begin{aligned}
HC_{2A} &= (P_{p1max} - (P_{REF} - 1))\,.\,D + (S_{SB} - 1)\,.\,D, where\ D \\
&\geq 2(T/R + 1)\ from\ Expression\ 1\ HC_{2A} \\
&= (P_{p1max} - P_{REF} + S_{SB})\,.\,D, where\ D \\
&\geq 2(T/R + 1) \ldots\ldots\ldots\ldots\ldots\ldots\ldots\ldots\ldots\ldots (Expression\ 15A)
\end{aligned}
$$

*Expression 15A* shows that the total hammer count a subbank can experience in Phase 2 until all preventive refreshes are produced, depends on its $PENDING$ value ($P_{REF}$) as well in addition to subbank size, blast radius, and parameter $D$. This is intuitive because the attacker can hammer the target subbank without experiencing any preventive refreshes until the target subbank's $PENDING$ value reaches $P_{p1max}$ in Phase 2-A.

[9] This prioritization scheme delays Silver Bullet from performing preventive refreshes on the target subbank as long as possible, enabling an attacker to perform more hammers on the target subbank.

### 5.2.3A Calculating the Total Hammer Count in Phase 2

We calculate the total hammer count that an attacker can achieve in Phase 2 as the sum of $HC_{2A}$ and $HC_{2B}$ (shown in Expression 17A).

$HC_2 = HC_{2A} + HC_{2B}, where$

- $HC_{2A} = (P_{p1max} - P_{REF} + S_{SB})\ D \qquad from\ Expression\ 15A$
- $HC_{2B} = T, \qquad from\ Expression\ 16$

$$HC_2 = (P_{p1max} - P_{REF} + S_{SB})\ D + T \ldots\ldots\ldots\ldots (Expression\ 17A)$$

## 5.3A Calculating the Attack's Hammer Count in the Worst Case

We calculate the total hammer count $(HC_{attack})$ an attacker can reach during Phases 1 and 2 by summing up the hammer counts of both phases $(HC_1 + HC_2)$.

$$HC_1 = D.P_{REF}, \quad if\ 1 \leq P_{REF} \leq P_{p1max} \qquad from\ Expression\ 13$$

$$\boldsymbol{P_{p1max} = log_2(N_{SB}) \qquad from\ Expression\ 11}$$

$$HC_2 = (\boldsymbol{log_2(N_{SB})} - P_{REF} + S_{SB})\ D + T \qquad from\ Expression\ 17A$$

Then, $HC_{attack} = HC_1 + HC_2$

$$HC_{attack} = (\boldsymbol{log_2(N_{SB})} + P_{REF} - P_{REF} + S_{SB})\ .D + T$$

$$HC_{attack} = (\boldsymbol{log_2(N_{SB})} + S_{SB})\ .D + T \ldots\ldots\ldots\ldots (Expression\ 18A)$$

## 5.4A Key Takeaways

In this section, we have calculated the total hammer count that an attacker can achieve if the attacker targets a subbank that was refreshed in Phase 1. Due to our worst-case assumption that there exists an infinite number of subbanks with a $PENDING$ value of $P_{p1max}$ in Phase 2, we calculate the *same* conservative upper-bound value for both this case and the case that the attacker targets a subbank that was never refreshed in Phase 1 (Section 5.3). Therefore, we can consider either of these cases throughout the remainder of the proof - we chose the latter in the main document.

# APPENDIX B. Security Analysis of the Extended Preventive Refresh Region Scheme

Silver Bullet proposes two different schemes of extending a subbank to account for cross-subbank RowHammer attacks. We already analyzed the method of extending the counter region throughout the main document. In this section, we analyze the second scheme: extending the preventive refresh region of a subbank. Figure 1B depicts the extended preventive refresh region.

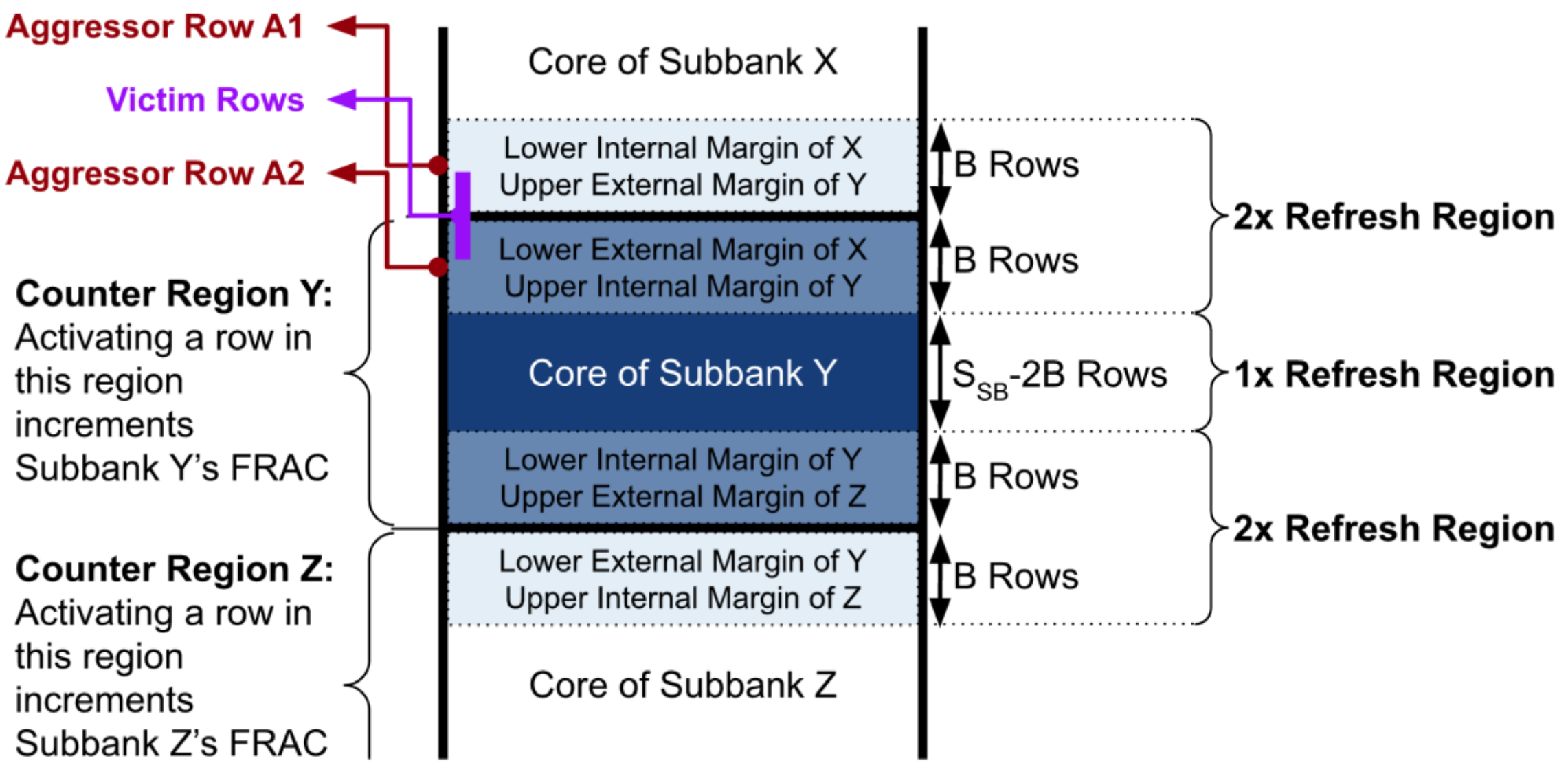

*Figure 1B. Extended Preventive Refresh Region Scheme*

In this scheme, counter regions of different subbanks are completely separated. For example in Figure 1B, the FRAC counter of Subbanks Y or Subbank Z is incremented only when a row within Subbank Y or within Subbank Z is activated, respectively. Therefore, there is no DRAM row that appears in two different subbanks' counter regions. As a result, a row activation increments the FRAC counter of only one subbank, as opposed to the extended counter region where a DRAM row can appear in two subbanks' counter regions. As such, one preventive refresh is produced for every D hammer, which includes both row activations and preventive refreshes. Therefore, the production/consumption rate criterion in Observation 1 is satisfied if the inequality in Expression 1B is satisfied.

$$1/D \leq R/(T+R). Then, D \geq (T+R)/R \ldots\ldots\ldots\ldots\ldots\ldots (Expression\ 1B)$$

## 5.1B Phase 1: Maximizing the Highest PENDING Value Across Subbanks

Assuming the usage of the extended refresh region scheme to address cross-subbank RowHammer effects, we update the attack model presented in Section 5.1.1 since the attacker can only increment the FRAC counter of at most one subbank with a single hammer. Therefore, we update Expressions 2 and 3 as shown in Expressions 2B and 3B, respectively.

$$N_{ACT}(i) = N(i) + (D-1)N(i+1) \ldots\ldots\ldots\ldots\ldots (Expression\ 2B)$$

$$N'(i) = \frac{R}{T}(N(i) + (D-1)N(i+1))\ldots\ldots\ldots\ldots\ldots\ldots\ldots (Expression\ 3B)$$

Following the same logical flow as Section 5 with updated Expressions 2B and 3B, we update Expression 5 as shown in Expression 5B.

$$N(i+1) \leq k\ N(i), where\ k = \frac{T-R}{T+(D-1)R}\ldots\ldots\ldots\ldots\ldots\ldots\ldots (Expression\ 5B)$$

Using Expressions 1B and 5B, we follow the same steps to calculate the maximum value that the subbank count reduction factor $(k)$ can take in Expression 6B.

$$k = \frac{T-R}{T+(D-1)\ R} = \frac{1-R/T}{1+(D-1)R/T}\ from\ Expression\ 5B$$

We solve this equation for $R/T$:

$$R/T = (1-k)/(kD-k+1)$$

$$Given\ that\ 1/D\ \leq R/T\ from\ Expression\ 1B:$$

$$1/D\ \leq\ (1-k)/(kD-k+1)$$

We solve this inequality for $k$:

$$k \leq (D-1)/(2D-1)\ldots\ldots\ldots\ldots\ldots\ldots\ldots\ldots\ldots\ldots\ldots\ldots\ldots\ldots\ldots\ldots (Expression\ 6B)$$

Based on Expression 6B, we define the same theoretical upper bound for k as in Expression 7. Therefore, we use Expression 8 as is, and we derive the same upper bound for $P_{p1max}$ as in Expression 11. We calculate the total hammer count the attacker can achieve in Phase 1 as we show in Expression 14B:

$$HC_1 = D\ .P_{p1max}, where\ D \geq T/R\ (from\ Expression\ 1B)\ldots\ldots\ldots\ldots\ldots (Expression\ 14B)$$

## 5.2B Phase 2: Maximizing the Hammer Count

To account for cross-subbank RowHammer attacks, Silver Bullet must perform preventive refreshes to all $S_{SB}+2B$ rows in the subbank's extended preventive refresh region (i.e., all rows in Subbank Y, its upper external margin in Subbank X, and its lower external margin in Subbank Z) when targeting a subbank. Note that in this scheme Subbank Y's counters are oblivious to any row activations that are performed on the upper and lower external margins of Subbank Y. Consider a case in which the attacker chooses and concurrently hammers one row from the upper external (i.e., Aggressor Row A1) and upper internal margins (i.e., Aggressor Row A2) of Subbank Y as highlighted in Figure 1B. While the attacker can reach a total hammer count of HC, both Subbanks X and Y observe only HC/2 hammers each. To address this attack model, this implementation of Silver Bullet performs preventive refreshes on rows in internal and external margins at double the rate of the rows in the subbank core. Therefore, Silver Bullet performs one preventive refresh on each row in a subbank core $(i.e., S_{SB}-2B$ rows) and two preventive refreshes on each row in upper/lower internal/external margins of a subbank $(i.e., 4B$ rows) within a preventive refresh window. Thus, we calculate the total number of preventive refreshes that must be performed to secure a subbank as $(S_{SB} - 2B) + (4B).2$. Therefore, we update Expressions 15, 17, 18, 20, and 22 by substituting $S_{SB}$with $S_{SB} + 6B$, as shown below in Expressions 15B, 17B, 18B, 20B, and 22B, respectively.

$$HC_{2A} = D.(S_{SB} + 6B) \ldots\ldots (Expression\ 15B)$$

$$HC_2 = D.(S_{SB} + 6B) + T \ldots\ldots (Expression\ 17B)$$

$$\boldsymbol{HC_{attack}} = (\boldsymbol{log_2}(\boldsymbol{N_{SB}}) + \boldsymbol{S_{SB}} + 6B)\,.\boldsymbol{D} + \boldsymbol{T} \ldots\ldots (\textbf{Expression 18B})$$

$$\boldsymbol{HC_{total}} = (\boldsymbol{log_2}(\boldsymbol{N_{SB}}) + \boldsymbol{S_{SB}} + \boldsymbol{6B})\,.\boldsymbol{D} + \boldsymbol{T} + \boldsymbol{2B} \ldots\ldots (Expression\ 20B)$$

$$THC = D\,.(\boldsymbol{log_2}(\boldsymbol{N_{SB}}) + \boldsymbol{S_{SB}} + \boldsymbol{6B}) + T + \boldsymbol{2B} \ldots\ldots (Expression\ 22B)$$

To compare extended counter region and extended preventive refresh region schemes, we restate $THC$calculation and restrictions of both schemes side-by-side below.

| **Extended Counter Region Scheme** | **Extended Preventive Refresh Region Scheme** |
|---|---|
| $From\ Expression\ 1: D \geq 2(T/R + 1)$ | $From\ Expression\ 1B: D \geq (T + R)/R$ |
| $From\ Expression\ 22:$ | $From\ Expression\ 22B:$ |
| $THC = D\,.(\boldsymbol{log_2}(\boldsymbol{N_{SB}}) + \boldsymbol{S_{SB}}) + T + \boldsymbol{2B}$ | $THC = D\,.(\boldsymbol{log_2}(\boldsymbol{N_{SB}}) + \boldsymbol{S_{SB}} + \boldsymbol{6B}) + T + \boldsymbol{2B}$ |

By comparing two schemes, we make two conclusions:

First, the extended counter region scheme does not support values of the parameter D as low as the extended preventive refresh region scheme. This limits the extended counter region scheme from securing DRAM chips that are extremely vulnerable to RowHammer. Note that reducing parameter D leads Silver Bullet to more aggressively producing preventive refreshes, which can allow extending security guarantees at the cost of potentially larger performance overheads.

Second, the extended preventive refresh scheme significantly increases the preventive refresh window of a subbank by increasing the subbank size by 6B compared to the extended counter region scheme. The effect of 6B additional preventive refreshes can be small in Silver Bullet implementations that employ large subbanks. However, our analyses in Section 10.2 shows that increasing the subbank size too much is undesirable due to larger resulting THC values.

## APPENDIX C. Leveraging Subarrays for Refresh Parallelism

A DRAM bank consists of groups of multiple DRAM rows, called subarrays [16,17] (or mid-banks). Subarrays are separated from each other by intermediary sense amplifier arrays called local sense amplifiers (or local row buffer). Local sense amplifiers physically isolate two adjacent subarrays from each other such that hammering a row in one subarray does not induce RowHammer bit flips in another subarray. Along with local sense amplifiers, each subarray has its own local row decoder and precharge circuitry, thereby potentially allowing a subarray to refresh a row in parallel with other subarrays. We assume that modern DRAM chips with RowHammer mitigation capability already exploit this subarray-level parallelism to be able to perform victim row refreshes during a periodic refresh command [18]. Thus, given a sufficient power budget, Silver Bullet can exploit this available parallelism to perform preventive refreshes.

Figure C1.a demonstrates an example set of rows that can be refreshed in parallel for an example set of four subarrays in a DRAM bank, each comprising of 8K rows such that:

- Subarray 0 contains the rows 0x0000 - 0x1FFF
- Subarray 1 contains the rows 0x2000 - 0x3FFF
- Subarray 2 contains the rows 0x4000 - 0x5FFF
- Subarray 3 contains the rows 0x6000 - 0x7FFF

Such a design can refresh 4x the rows that a bank without subarray-level parallelism refreshes during a periodic refresh command, which is called over-refreshing. Silver Bullet can leverage a portion of the over-refreshing capability (e.g., 1/4 or 2/4 of refreshes) to perform preventive refreshes in the presence of periodic refreshes. Thus, a significant factor of periodic over-refreshing can still be maintained (e.g., 3x or 2x) for the cells that require over-refreshing, regardless of RowHammer considerations.

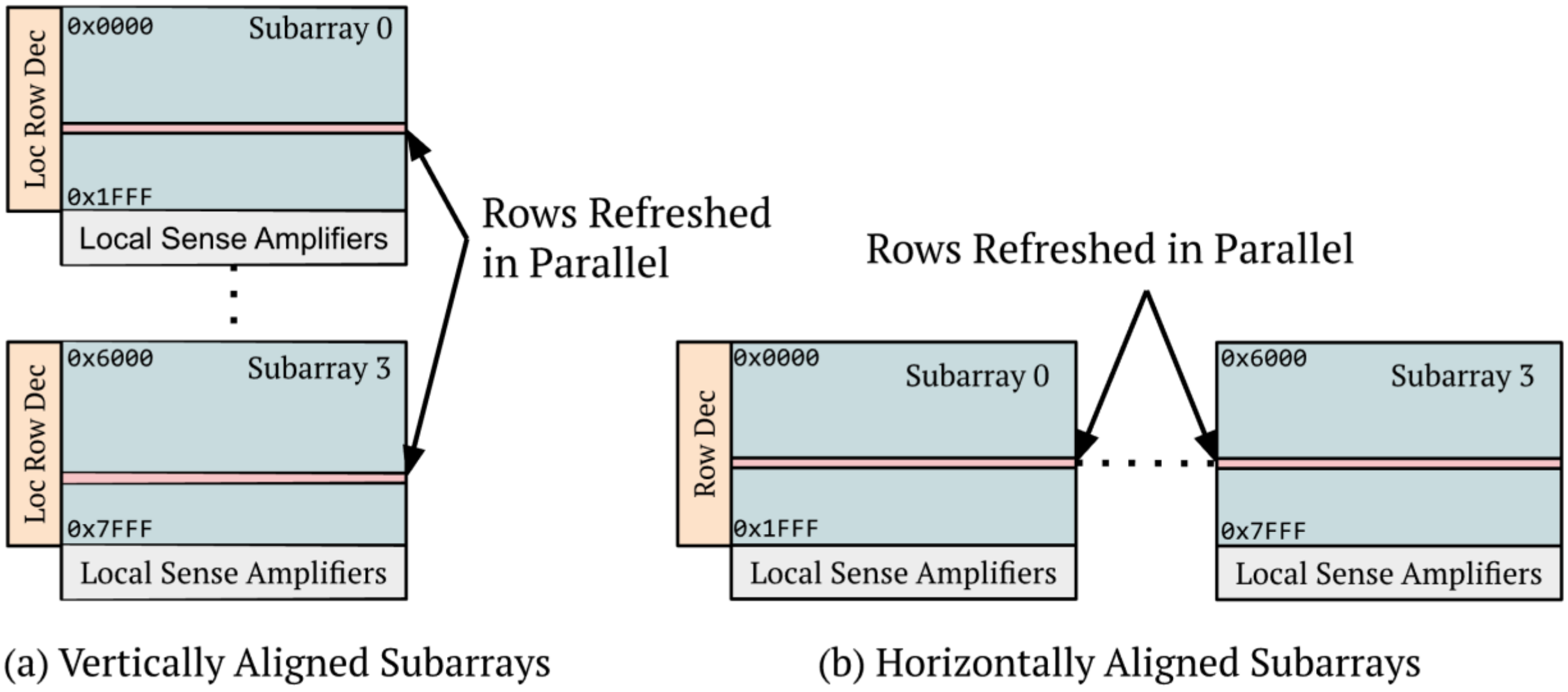

*Figure C1. Exploiting Subarray-level Parallelism for Refresh*

As a design optimization, these four subarrays can be aligned horizontally to share a row decoder as shown in Figure C1.b such that for a given row address N, subarrays 0, 1, 2, and 3 simultaneously refresh rows N, 0x4000+N, 0x8000+N, and 0xC000+N, respectively. With such a design optimization, each preventive or periodic refresh refreshes four rows in four different subarrays in parallel, enabling Silver Bullet to (1) perform multiple preventive refreshes within the time dedicated for periodic refreshes and (2) reduce its table size by maintaining only one table per grouping of subarrays that are refreshed together in parallel.

Note that over-refreshing *cannot prevent* RowHammer bit flips but *can mitigate* the RowHammer effect by shortening the refresh window and therefore reducing the time period available to an attacker.